\RequirePackage{lineno}
\documentclass[aps,prl,amssymb,12pt,preprint,nofootinbib,floatfix]{revtex4-1}

\usepackage{hyperref,graphicx}
\usepackage{verbatim}
\usepackage{subfig}

\begin{document}
\def\aj{\rmfamily{AJ}}%
\def\actaa{\rmfamily{Acta Astron.}}%
\def\araa{\rmfamily{ARA\&A}}%
\def\apj{\rmfamily{ApJ}}%
\def\apjl{\rmfamily{ApJ}}%
\def\apjs{\rmfamily{ApJS}}%
\def\ao{\rmfamily{Appl.~Opt.}}%
\def\apss{\rmfamily{Ap\&SS}}%
\def\aap{\rmfamily{A\&A}}%
\def\aapr{\rmfamily{A\&A~Rev.}}%
\def\aaps{\rmfamily{A\&AS}}%
\def\azh{\rmfamily{AZh}}%
\def\baas{\rmfamily{BAAS}}%
\def\caa{\rmfamily{Chinese Astron. Astrophys.}}%
\def\cjaa{\rmfamily{Chinese J. Astron. Astrophys.}}%
\def\icarus{\rmfamily{Icarus}}%
\def\jcap{\rmfamily{J. Cosmology Astropart. Phys.}}%
\def\jrasc{\rmfamily{JRASC}}%
\def\memras{\rmfamily{MmRAS}}%
\def\mnras{\rmfamily{MNRAS}}%
\def\na{\rmfamily{New A}}%
\def\nar{\rmfamily{New A Rev.}}%
\def\pra{\rmfamily{Phys.~Rev.~A}}%
\def\prb{\rmfamily{Phys.~Rev.~B}}%
\def\prc{\rmfamily{Phys.~Rev.~C}}%
\def\prd{\rmfamily{Phys.~Rev.~D}}%
\def\pre{\rmfamily{Phys.~Rev.~E}}%
\def\prl{\rmfamily{Phys.~Rev.~Lett.}}%
\def\pasa{\rmfamily{PASA}}%
\def\pasp{\rmfamily{PASP}}%
\def\pasj{\rmfamily{PASJ}}%
\def\qjras{\rmfamily{QJRAS}}%
\def\rmxaa{\rmfamily{Rev. Mexicana Astron. Astrofis.}}%
\def\skytel{\rmfamily{S\&T}}%
\def\solphys{\rmfamily{Sol.~Phys.}}%
\def\sovast{\rmfamily{Soviet~Ast.}}%
\def\ssr{\rmfamily{Space~Sci.~Rev.}}%
\def\zap{\rmfamily{ZAp}}%
\def\nat{\rmfamily{Nature}}%
\def\iaucirc{\rmfamily{IAU~Circ.}}%
\def\aplett{\rmfamily{Astrophys.~Lett.}}%
\def\apspr{\rmfamily{Astrophys.~Space~Phys.~Res.}}%
\def\bain{\rmfamily{Bull.~Astron.~Inst.~Netherlands}}%
\def\fcp{\rmfamily{Fund.~Cosmic~Phys.}}%
\def\gca{\rmfamily{Geochim.~Cosmochim.~Acta}}%
\def\grl{\rmfamily{Geophys.~Res.~Lett.}}%
\def\jcp{\rmfamily{J.~Chem.~Phys.}}%
\def\jgr{\rmfamily{J.~Geophys.~Res.}}%
\def\jqsrt{\rmfamily{J.~Quant.~Spec.~Radiat.~Transf.}}%
\def\memsai{\rmfamily{Mem.~Soc.~Astron.~Italiana}}%
\def\nphysa{\rmfamily{Nucl.~Phys.~A}}%
\def\physrep{\rmfamily{Phys.~Rep.}}%
\def\physscr{\rmfamily{Phys.~Scr}}%
\def\planss{\rmfamily{Planet.~Space~Sci.}}%
\def\procspie{\rmfamily{Proc.~SPIE}}%
\let\astap=\aap
\let\apjlett=\apjl
\let\apjsupp=\apjs
\let\applopt=\ao

\renewcommand{\deg}
{\ensuremath{^{\circ}}}

\title{A Method for Constraining Cosmic Magnetic Field Models Using Ultra-High Energy Cosmic Rays: The Field Scan Method}

\author{Michael S.~Sutherland}%
\email[Author's email: ]{msutherland@phys.lsu.edu}%
\affiliation{
  Department of Physics and Astronomy,
  Louisiana State University,
  Baton Rouge, LA 70803, USA
  }%
\author{Brian M.~Baughman}%
\email[Author's email: ]{bbaugh@mps.ohio-state.edu}%
\affiliation{
  Department of Physics,
  University of Maryland,
  College Park, MD 20742, USA
  }%
\altaffiliation{
  Department of Physics and the Center for Cosmology and Astro-Particle Physics,
  The Ohio State University,
  Columbus, Ohio 43210, USA
}%
\author{J. J.~Beatty}%
\email[Author's email: ]{beatty@mps.ohio-state.edu}%
\affiliation{
  Department of Physics and the Center for Cosmology and Astro-Particle Physics,
         The Ohio State University,
         Columbus, Ohio 43210, USA
}%

\begin{abstract}
The Galactic magnetic field, locally observed to be on the order of a few $\mu$G, is sufficiently strong to induce deflections in the arrival directions of ultra-high energy cosmic rays.
We present a method that establishes measures of self-consistency for hypothesis sets comprised of cosmic magnetic field models and ultra-high energy cosmic ray composition and source distributions.
The method uses two independent procedures to compare the backtracked velocity vectors outside the magnetic field model to the distribution of backtracked velocity directions of many isotropic observations with the same primary energies.
This allows for an estimate of the statistical consistency between the observed data and simulated isotropic observations.
Inconsistency with the isotropic expectation of source correlation in both procedures is interpreted as the hypothesis set providing a self-consistent description of GMF and UHECR properties for the cosmic ray observations.
\end{abstract}
\maketitle

\section{Introduction}
\label{sec:introduction}
Ultra-high energy cosmic rays (UHECRs) are almost certainly extragalactic in origin as no anisotropy is observed except at the highest energies \cite{2007Sci...318..938T}.
UHECRs are also thought to be largely comprised of charged particles \cite{2009APh....31..399T} and will therefore experience magnetic deflection during propagation from their sources.
This deflection can be sufficiently large as to make source identification difficult.

The UHECR magnetic deflection is thought to arise as two distinct contributions: extragalactic and Galactic.
Turbulent magnetic fields are expected to comprise the dominant component in extragalactic space with nG field strength upper limits \cite{1994RPPh...57..325K, 1998AIPC..433..196K, 2001SSRv...99..243B, 2006AN....327..517K, 2008AIPC.1085...83B}.
There is disagreement regarding the expected typical deflection, ranging from less than a few degrees to tens of degrees (see, e.g., \cite{1998A&A...335...19R, 2004PhRvD..70d3007S, 2004JETPL..79..583D, 2005JCAP...01..009D, 2008JPhCS.120f2025D, 2010ApJ...710.1422R}).
However, these studies are highly sensitive to simulation conditions, such as propagation through filaments and voids with strong and weak fields, respectively.

The deflection induced by the Galactic magnetic field (GMF) is thought to dominate over the extragalactic deflection (see, e.g., \cite{1997ApJ...479..290S, 1998ApJ...492..200M, 1999JHEP...08..022H, 2000JHEP...02..035H, 2006ApJ...639..803T, 2007APh....26..378K, 2008ApJ...681.1279T, 2010ApJ...710.1422R}), owing to the significantly larger field strengths and more organized (regular) field structure.
Simulated isotropically observed cosmic rays are backtracked through particular GMF configurations to determine the typical deflection magnitudes along different sight-lines.
Deflection magnitudes at least of order few degrees are common, even for proton primaries with energies greater than a few $10^{18}$ eV.

However, the GMF structure is not well-known aside from the neighborhood of the Sun.
Near the Sun, the local field is observed to point towards Galactic longitude $\ell \approx 80\deg$ in the Galactic plane \cite{1970MmRAS..74..139M, 1996ApJ...462..316H, 2000AJ....119..923H} following the stellar spiral arm with a regular component field strength of order few $\mu$G \cite{1994MNRAS.268..497R, 2002ApJ...570L..17H, 2004Ap&SS.289..293B, 2007ApJ...663..258B, 2009Natur.462.1036O}.
The turbulent component is thought to be of roughly the same magnitude with typical cell sizes less than 100 pc \cite{1971A&A....14..359B, 1976A&AS...26..129B, 1996ASPC...97..457H, 2004Ap&SS.289..293B, 2010ApJ...714.1170M}.
Larger field strengths are observed closer to and within the Galactic center (see, e.g., \cite{2010Natur.463...65C}).
A regular component extends to a distance of at least 500 pc from Galactic plane (see, e.g., \cite{2001A&A...368..635B}).
Galactic halo fields may extend well beyond this according to expectations with the Galactic electron distribution (see, e.g., \cite{2010RAA....10.1287S}).
There is additionally disagreement regarding the number and location of field reversals within the disk, although there is strong evidence of at least one reversal inside the solar circle and possibly none outside (see, e.g., \cite{1994MNRAS.268..497R, 2006ApJ...642..868H, 2007ApJ...663..258B, 2011ApJ...728...97V}).

There is currently no consensus on the structure of the large-scale regular component of the GMF \cite{2004Ap&SS.289..293B, 2008A&A...486..819M, 2009JCAP...07..021J}.
Observations of magnetic fields in similar galaxies may provide clues regarding the magnetic structure of the Galaxy.
Spiral galaxies are typically observed to possess a regular magnetic field component following the spiral arms \cite{1994RPPh...57..325K, 2001SSRv...99..243B, 2004NewAR..48..763V}.
In the interarm regions, the regular magnetic field dominates over the turbulent component, and vice-versa within the arms \cite{2000A&A...357..129E, 2008A&A...477..573S, 2008A&A...490.1005T, 2009Ap&SS.320...77B}.
Observed field strength magnitudes are typically of order $1-25\ \mu$G within the disk \cite{2001SSRv...99..243B, 2004NewAR..48..763V, 2004Ap&SS.289..293B}.
However, numerous starburst galaxies have been observed with magnetic field strengths of up to $100\mu$G \cite{1988A&A...190...41K, 2004A&A...417..541C, 2005A&A...444..739B, 2008Natur.455..638W}.

Assuming an extragalactic source distribution, one could backtrack observed cosmic rays through various field models until a significant number coincided with hypothesized sources.
However, a large number of coincident events may simply be a feature of a field model despite knowing the UHECR composition and source distribution.
Harari et al. \cite{2000JHEP...02..035H} have shown that certain GMF models exhibit lensing that focuses cosmic ray trajectories onto specific parts of the extragalactic sky.
This effect may artificially increase the number of coincident correlations with hypothesized sources located in regions where lensing is strong.
For example, the report \cite{2002APh....18..165T} of a strong correlation between the backtracked AGASA UHECR dataset and selected BL Lac objects was found to be highly dependent on the composition and GMF hypotheses \cite{2007APh....26..378K}.

In this paper a method for determining self-consistency of cosmic magnetic field model and cosmic ray property hypothesis sets for UHECR datasets is presented, henceforth referred to as the Field Scan Method (FSM).
The cosmic ray dataset and many isotropic simulations are backtracked to determine the extragalactic arrival distribution.
An isotropic arrival distribution of cosmic rays at Earth will propagate back through the magnetic field and map into a non-isotropic extragalactic arrival distribution due to magnetic lensing.
A highly anisotropic sky distribution and an excess of coincident events with the hypothesized source distribution compared to the expectation derived from the isotropic simulations serve as indications that the magnetic field model together with the composition and source distribution hypotheses provide a self-consistent description of the cosmic ray observations.
The comparisons against isotropy allow for the hypothesis consistency test even in a scenario where magnetic lensing is strong along particular lines of sight.
Preliminary implementations of this method were explored in \cite{phdthesis-full, 2009arXiv0906.2347T}, however, in this paper we optimize the analysis with respect to the previously applied methods.
Table \ref{tbl:acronyms} contains a reference list of acronyms used throughout the text.
\begin{table}[htp]
\begin{center}
\caption{Table of Acronyms}
\begin{tabular}{|c|c|}
 \hline
 DOI & Dataset of interest \\
 \hline
 MCISOs & Monte Carlo simulations of isotropy \\
 \hline
 $N_{corr}^{DOI}$, $\overline{N}_{corr}^{iso}$ & Number of correlating events in the dataset \\
 \hline
 $\overline{N}_{corr}^{iso}$ & Mean number of correlating events in the isotropic simulations \\
 \hline
 $P_{bin}$ & Probability of drawing $N_{corr}^{DOI}$ correlating events from a binomial distribution \\
           & defined by the correlation properties of the isotropic simulations \\
 \hline
 $D_{max}$, $P_{KS}$ & Maximum signed difference between the dataset and isotropic test statistic \\
                     & distributions and the probability that the distributions result from the \\
                     & same parent distribution \\
 \hline
 $\Theta_{sep}^{min}$ & Normalized minimum angular separation between the magnetically- \\
                       & corrected arrival direction and the assigned ``nearest'' source object \\
 \hline
 $\delta_{B}$ & Angular deflection magnitude of an event \\
 \hline
 $\Psi$ & Test statistic \\
 \hline
 $\theta_{src}$ & Actual angular distance between an event and a source object \\
 \hline
 $\delta_{turb}$ & RMS angular deflection from a turbulent magnetic field \\
 \hline
 RADIO & Catalog of radio galaxies \\
 \hline
 RADIO+SWIFT39 & Catalog of radio galaxies and selected objects in the Swift 39-month catalog \\
 \hline
 ISOTROPIC80 & List of 80 isotropically selected directions \\
 \hline
 VCV & 12$^{th}$ Edition V\'{e}ron-Cetty and V\'{e}ron catalog \\
 \hline
 BSS$\_$A, ASS$\_$S & Two variation of logarithmic sprial GMF models \\
 \hline
\end{tabular}
\label{tbl:acronyms}
\end{center}
\end{table}

\section{Procedure}
\label{sec:fieldscan-procedure}
\begin{figure}
\includegraphics[width=0.7\textwidth]{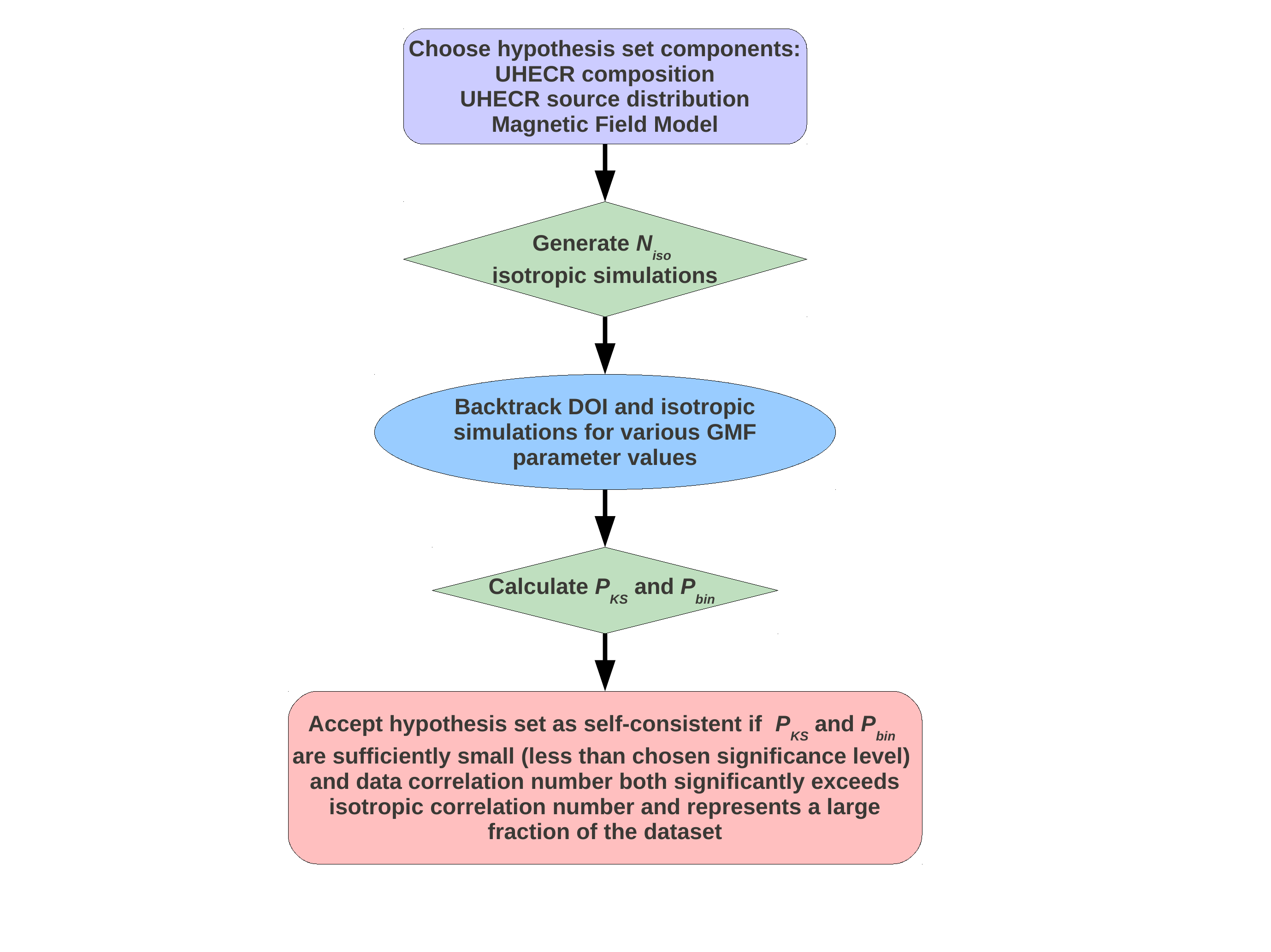}
\caption{Methodology of the Field Scan Method.}
\label{fig:flowchart}
\end{figure}

The FSM provides an estimate of hypothesis set self-consistency by comparing the behavior of an UHECR dataset of interest (DOI) folded with the hypothesis set against that of Monte Carlo simulations of isotropy (MCISOs).
Figure \ref{fig:flowchart} shows the general procedure of the method.
Hypothesis sets contain three independent components: observed composition, source distribution\footnote{Any catalog which correlates well with the true source distribution will show a deviation from the correlation expectation from isotropy.}, and a magnetic field configuration.
The MCISOs are constructed with the same DOI energies, but with directions sampled from a homogeneous isotropic flux, in the vicinity of where the observations were taken, modulated by detector coverage matching that of the DOI.
The method is strongly dependent on the underlying hypothesis set components.
The correctness of individual hypothesis components cannot be determined, only that the set is self-consistent in its entirety.

The essential idea behind the method is to find regions of the hypothesis set parameter space\footnote{That is, any parameters which are not otherwise constrained.} for which source correlation and consistency with the isotropic simulations are maximized and minimized respectively.
Such regions of parameter space cannot be ruled out as self-consistent theories describing the observed composition, source distribution, and magnetic field, therefore meriting further examination.
Regions of parameter space where the DOI is found to be more consistent with the MCISOs or exhibit unidentifiably significant source correlation represent hypotheses which are not self-consistent and can thus be ruled out as viable theories. 

Self-consistency is determined by two independent procedures: (1) correlation using angular windows and (2) a statistical test performed on the distribution of an event-by-event test statistic (TS) for the DOI and MCISOs.
In procedure (1), a correlation number is determined by counting the events which fall within a specified angular window around the source objects after magnetic correction.
The DOI correlation number is then compared to the correlation number distribution from the isotropic simulations.
For small angular windows the isotropic correlation number distribution closely matches a binomial distribution, so that the binomial probability $P_{bin}$ of obtaining the DOI correlation value can be computed from,
\begin{equation}
\label{eqn:binomial}
P_{bin}(k;n,p)={n \choose k}p^{k}(1-p)^{n-k}
\end{equation}
The probability of success $p$ can be approximated by ($\overline{N}^{iso}_{corr}$ / $N_{tot}$), where $\overline{N}^{iso}_{corr}$ is the mean number of correlating events from many isotropic instances and $N_{tot}$ is the total number of events in the DOI or a single isotropic instance.
$P_{bin}$ can be calculated using $k=N^{DOI}_{corr}$ and $n=N_{tot}$.
A large positive excess of correlating events will yield a small probability, indicating that the hypothesis set is self-consistent.
A negative or small positive excess indicates that the DOI does not correlate more than the isotropic expectation and does not indicate identifiable self-consistency.
Magnetic lensing onto a portion of the sky containing many source objects may present a scenario where $N^{DOI}_{corr}$ itself is large but the excess small.

Procedure (2) is designed to identify those hypothesis sets that do not produce identifiably significant source correlation.
It determines the compatibility between distributions of event-by-event TS values for the DOI and MCISOs.
After backtracking the $i^{\rm th}$ event using the hypothesized composition and magnetic field model the test statistic,
\begin{equation}
\label{eqn:psi}
\Psi_{i} = \frac{\Theta_{sep,i}^{min}}{1.+\delta_{B,i}}
\end{equation}
is calculated where $i$ denotes the $i^{\rm th}$ event, $\delta_{B,i}$ is the angular deflection magnitude between the observed and magnetically-corrected arrival vectors, and $\Theta_{sep,i}^{min}$ is the normalized minimum angular separation between the magnetically-corrected arrival vector and the assigned ``nearest'' source object.
This construction of $\Psi_{i}$ does not diverge as the magnitude of the magnetic field goes to zero ($\delta_{B,i} \to 0$) and is minimized when $\Theta_{sep,i}^{min}$ is also minimized (for monotonic $\delta_{B,i}$).
A normalized minimum angular separation can incorporate an estimate of deflection through a turbulent magnetic field, e.g., an extragalactic magnetic field.
Such fields have been shown to induce a gaussian smearing effect on the velocity vectors of UHECRs injected into them \cite{2004IJMPA..19.1133R}.
For example, the assigned ``nearest'' source object could be that which minimizes the ratio $\theta_{src}/\delta_{turb}$, where $\theta_{src}$ is the actual angular distance between the backtracked event and a source object and $\delta_{turb}$ is the estimate of RMS magnetic deflection from the source object induced by a fixed parameter turbulent field.
The RMS deflection magnitude would depend on the characteristic magnitude and correlation length of the turbulent model, as well as the event energy and source distance.
In such a case $\Theta_{sep}^{min}=\mbox{min}(\theta_{src}/\delta_{turb})$.
In cases where no such turbulent field is modeled, $\Theta_{sep}^{min}$ would be simply $\mbox{min}(\theta_{src})$.
Use of a normalized minimum angular separation in general can allow for the use of a much deeper, in redshift, catalog as this normalization scales the correlation window with distance from the source.

The statistical compatibility of the full DOI with the MCISOs is determined using a Kolmogorov-Smirnov (KS) test\footnote{Other hypothesis testing algorithms can be easily substituted.} between the distributions of $\Psi^{\rm{DOI}}_{i}$ for the DOI and $\Psi^{\rm{MCISO}}_{j}$ for MCISOs.
This test employs the cumulative distribution function (CDF) of $\Psi_{i}$ distribution and determines the maximum signed difference $D_{max}$ of the quantity,
\begin{equation}
\label{eqn:ddist}
D_{i} = CDF^{\rm{DOI}}(\Psi_{i})
- CDF^{\rm{MCISO}}(\Psi_{i})
\end{equation}
between the $\Psi_{i}$ CDFs of the DOI and MCISO.
$D_{max}$ maps to the probability $P_{KS}$ that the distribution from which the $\Psi^{\rm{DOI}}_{i}$ values are drawn is similar to the distribution from which the $\Psi^{\rm{MCISO}}_{j}$ values are drawn.
As constructed in Eqn. \ref{eqn:psi}, a smaller $\Psi_{i}$ indicates stronger correlation between the hypothesized source distribution and the backtracked trajectory of the tested event.
By examining the sign and location of $D_{max}$, it is possible to determine if source correlations are increasing under the hypothesis set.
This ensures that the entire hypothesis set (composition, magnetic field model, and source distribution) is being tested and that the minimum in consistency between $\Psi^{\rm{DOI}}_{i}$ and $\Psi^{\rm{MCISO}}_{j}$ is indicating an increased correlation with the source distribution.

The hypothesis set is deemed self-consistent to the extent that the $P_{KS}$ value indicates inconsistency with isotropy and that the DOI correlates well with the source hypothesis.
A large positive $D_{max}$, resulting in small $P_{KS}$, located at a small TS value indicates that the DOI better correlates with the source hypothesis than the isotropic expectation and is inconsistent with the isotropic expectation.
Conversely, if the dataset differs little from the isotropic expectation (small $D_{max}$ and large $P_{KS}$), then one or more of the hypothesis components may be incorrect, or perhaps the method is probing a regime where self-consistency cannot be identified (e.g., strong lensing that hinders identification of significant source correlation beyond the isotropic expectation).
Positive $D_{max}$ at large TS values and any negative $D_{max}$ are also indicators of these scenarios.

A typical significance level for rejecting the null hypothesis (the DOI and isotropic TS distributions appear to be drawn from the same parent distribution) is $\alpha=1\%$, i.e., $P_{KS}<0.01$.
This value is appropriate for determining whether the DOI is similar to isotropy in this method.

These procedures provide complementary results.
Procedure (1) allows for a numerical correlation estimate as well as the probability of such result, but is insensitive to details of the backtracked sky distribution and the final locations of the events around the source objects.
Procedure (2) incorporates the source separation and magnetic deflection magnitude for every event to determine the overall consistency between the backtracked sky distributions for the DOI and MCISOs.
Significant deviations from the expectation of isotropic correlation are expected in both procedures under a hypothesis set that correctly describes the UHECR observations.


\section{Validation}
\label{sec:validation}
The FSM has been tested by simulating a variety of simple truth scenarios, as well as spanning more realistic configurations of magnetic fields and UHECR composition and source distributions.
Here we present the results of one such simple scenario; realistic scenarios are briefly described at the end of this section.

\begin{figure}
\includegraphics[width=0.7\textwidth]{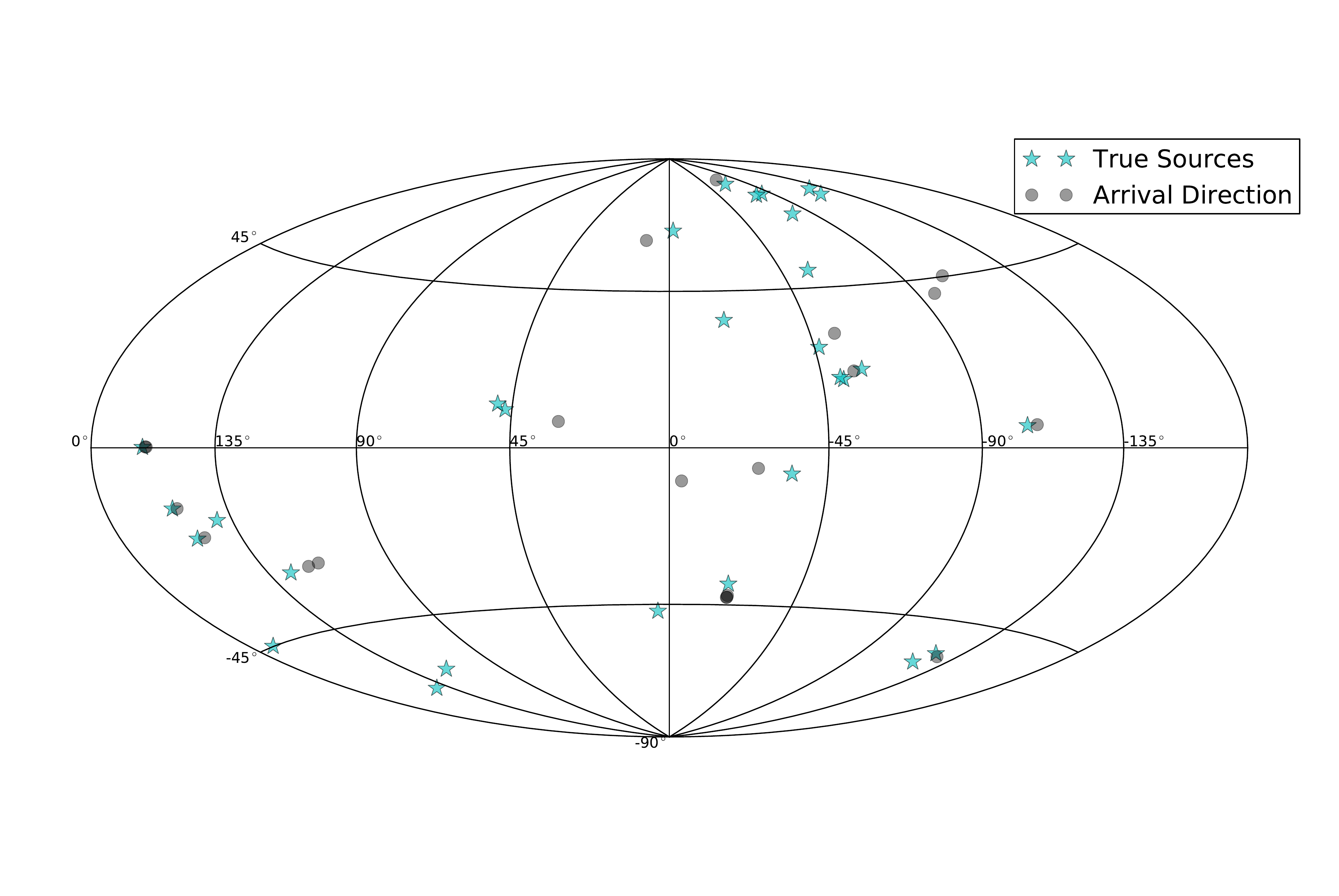}
\caption{Observed arrival directions (shaded black circles) and positions of their true sources (blue stars) of the 20 cosmic rays in the validation scenario.}
\label{fig:src_arrival}
\end{figure}

A mock universe is constructed using the parameters of Table \ref{tbl:mock-truth}.
20 cosmic ray events were generated such that, when backtracked through the truth magnetic field model configuration assuming the truth composition, they lie within $0.01\deg$ of an object from the truth source distribution.
The true source positions and the observed arrival directions are shown in Figure \ref{fig:src_arrival}.
The truth hypothesis set is tested to confirm that both methods correctly identify self-consistency at the correct field parameter values and that self-consistency decreases as the hypothesis set is altered.
Various non-truth hypothesis sets are also tested to determine the possibility of generating a false signal.
These simulations are backtracked using \textit{CRT}, a public numerical tool for propagating UHECRs through magnetic field model models \cite{Sutherland2010}.
\begin{table}[htp]
\begin{center}
\caption{Truth Scenario}
\begin{tabular}{|c|c|}
 \hline
 Composition                        &   Pure proton             \\
 \hline
 Source Distribution                &   RADIO            \\
 \hline
 GMF                         &   Pure Dipole           \\
 \hline
 Local Field Strength ($B_{\odot}$) &   $1.5\ \mu$G          \\
 \hline
\end{tabular}
\label{tbl:mock-truth}
\end{center}
\end{table}

\subsection{Composition Hypothesis Component}
\label{sec:validation:composition}
A pure proton composition is hypothesized for all combinations of source and GMF hypothesis components.

\subsection{GMF model component}
\label{sec:magneticfields}
The parameter space of two distinct models are scanned: pure dipole and pure uniform.
The pure dipole is the truth field.
In both models, the only parameter is the local field strength which is scanned from -1.0 $\mu$G to 3.0 $\mu$G in steps of 0.1 $\mu$G.
A positive dipole field strength gives a field vector oriented towards the North Galactic Pole.
The uniform field is oriented towards the Galactic longitude $\ell=90^{\circ}$ for positive field magnitudes, i.e. perpendicular to the direction of the Galactic center and wholly parallel to the Galactic plane.
Galactic turbulent fields are not modeled nor any turbulence in extragalactic space.
Both fields have zero magnitude beyond a galactocentric distance of 20 kpc.

\subsection{Source Catalogs}
\label{sec:validation:catalogs}
Three individual source distributions are tested: a selection of 29 radio galaxies \cite{2009NuPhS.190...61B} (RADIO) which also comprise the truth source distribution, a combination of RADIO and the Palermo Swift-BAT hard X-ray catalog \cite{2010A&A...510A..48C} (RADIO+SWIFT39), and a catalog comprised of 80 isotropically selected directions (ISOTROPIC80).
A redshift cut of $z_{max}\leq0.018$ is applied only to RADIO+SWIFT39 resulting in 127 objects (21 from RADIO and 106 from SWIFT39).

\subsection{Results}
\label{sec:validation:results}
\begin{figure}
\includegraphics[width=0.7\textwidth]{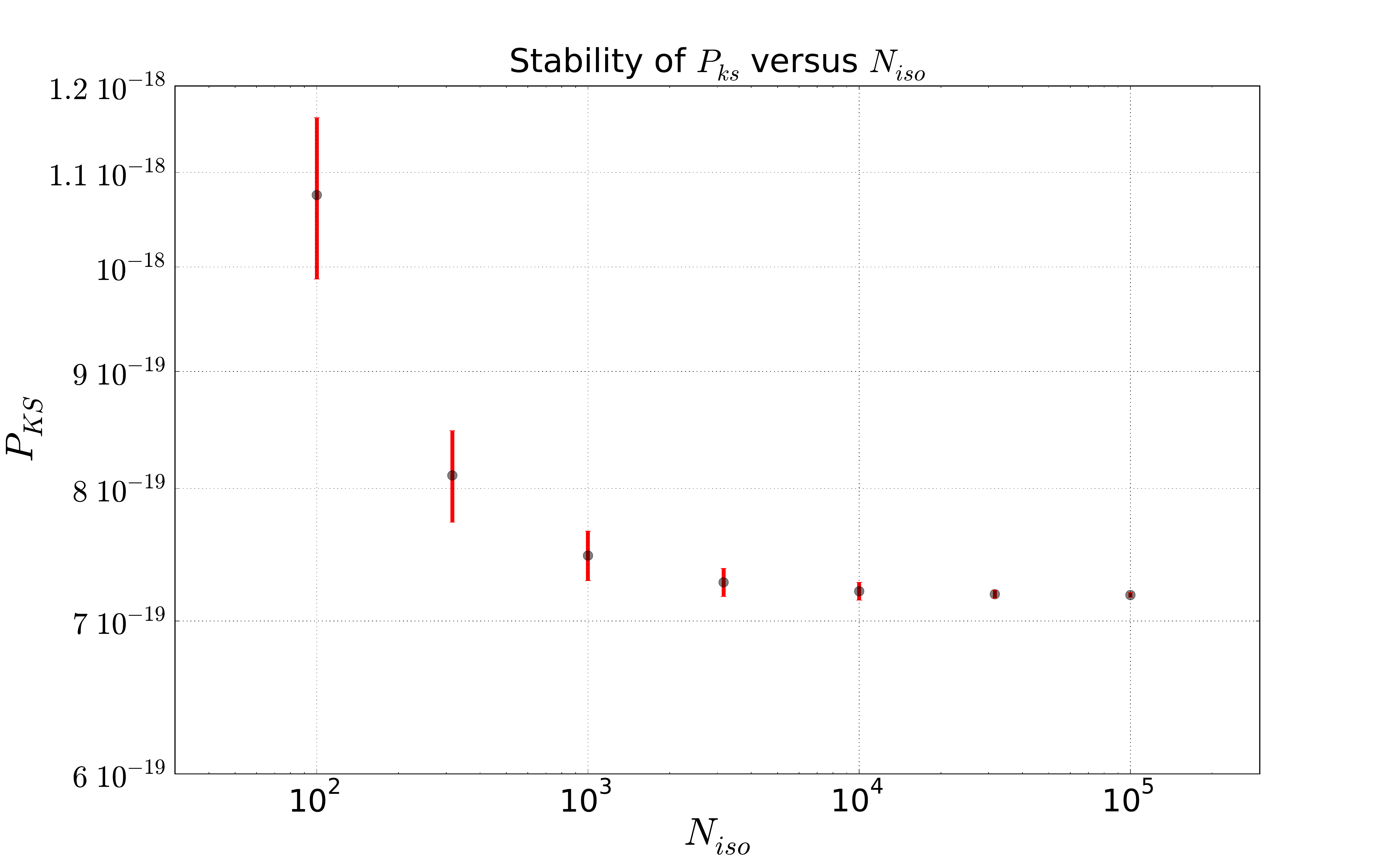}
\caption{Distribution of $P_{KS}$ for 100 simulations at different values of $N_{iso}$.
The red error bars depict the spread of the distribution at each point.}
\label{fig:varyNiso}
\end{figure}

We first perform a scan over the truth hypothesis set (Proton, RADIO, Dipole) at $B_{\odot}=1.5\ \mu$G by varying the number of isotropic instances $N_{iso}$ used to determine $P_{KS}$.
This will allow a determination of an appropriate $N_{iso}$ that balances accurate modeling of isotropy with computation time.
100 unique simulations are generated for each value of $N_{iso}$ in the range ($10^{2}$, $10^{2.5}$, ... , $10^{5}$).
The resulting distributions of $P_{KS}$ are shown in Figure \ref{fig:varyNiso}.
For large values of $N_{iso}$, the mean of the distribution approaches a constant value and the distribution itself tightens, as illustrated by the red error bars.
There appears to be no gain in accuracy or precision beyond $N_{iso}=10^{4}$ and only corresponding increases in computation time.

\begin{figure}
  \centering
  \subfloat[]{\label{fig:p-d-r_ncorr}\includegraphics[width=0.5\textwidth]{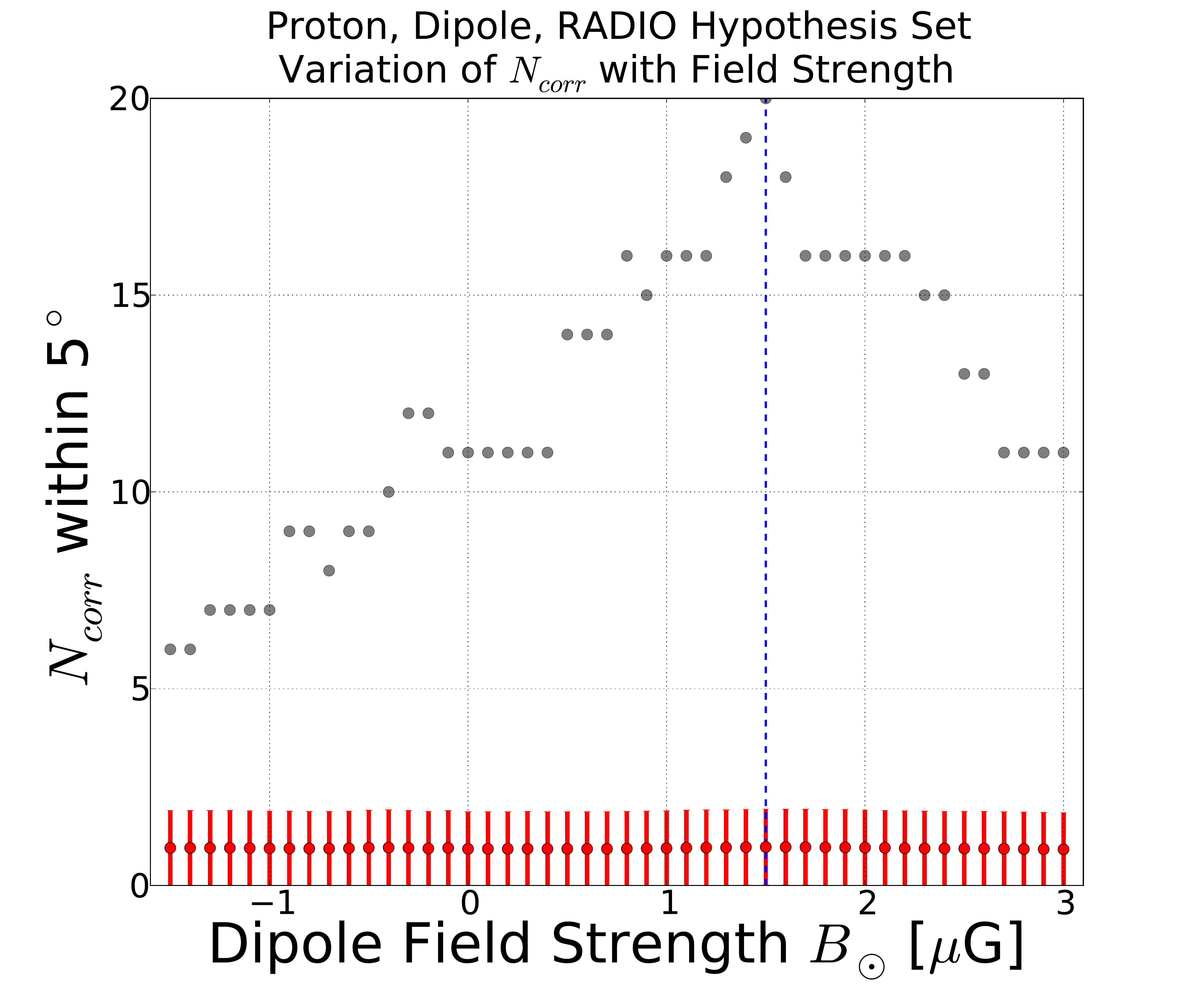}}
  \subfloat[]{\label{fig:p-d-r_Pks}\includegraphics[width=0.5\textwidth]{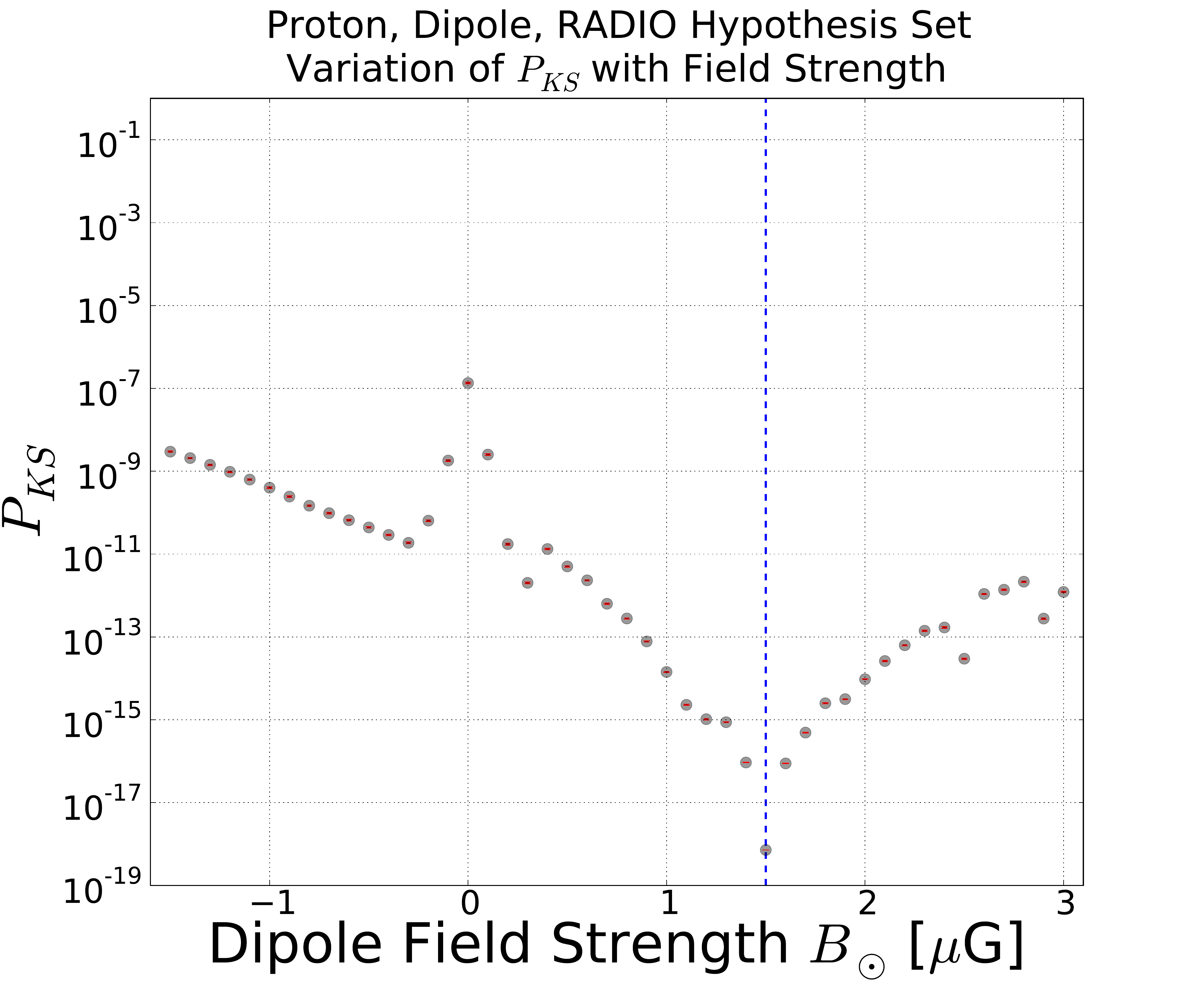}}
  \caption{Magnetic field strength scan of the truth scenario.
The truth field strength is indicated by the dashed blue line.
The left plot shows the number of events correlating with RADIO sources within $5^{\circ}$ windows.
The red dots and error bars indicate the mean and spread of the isotropic correlation number distributions.
The right plot shows the distribution of $P_{KS}$.
The red shaded dots and red lines indicate the mean and spread of the $P_{KS}$ distribution from the 100 isotropic simulations.}
  \label{fig:p-d-r}
\end{figure}

We now scan the magnetic field strength parameter space of the truth hypothesis set between $-1.5$ and $3.0\ \mu$G.
At each field strength point, we generate 100 simulations each comprised of $N_{iso}=10^{4}$ unique isotropic instances.
Figure \ref{fig:p-d-r_ncorr} indicates that the mean isotropic correlation number is essentially constant across all scanned field strengths, whereas better correlation in the DOI is observed closer to the correct magnetic field arrangement.
A very clear minimum in $P_{KS}$ is also observed in coincidence with the highest source correlation at the correct field strength value in Figure \ref{fig:p-d-r_Pks}.

\begin{figure}
\includegraphics[width=0.7\textwidth]{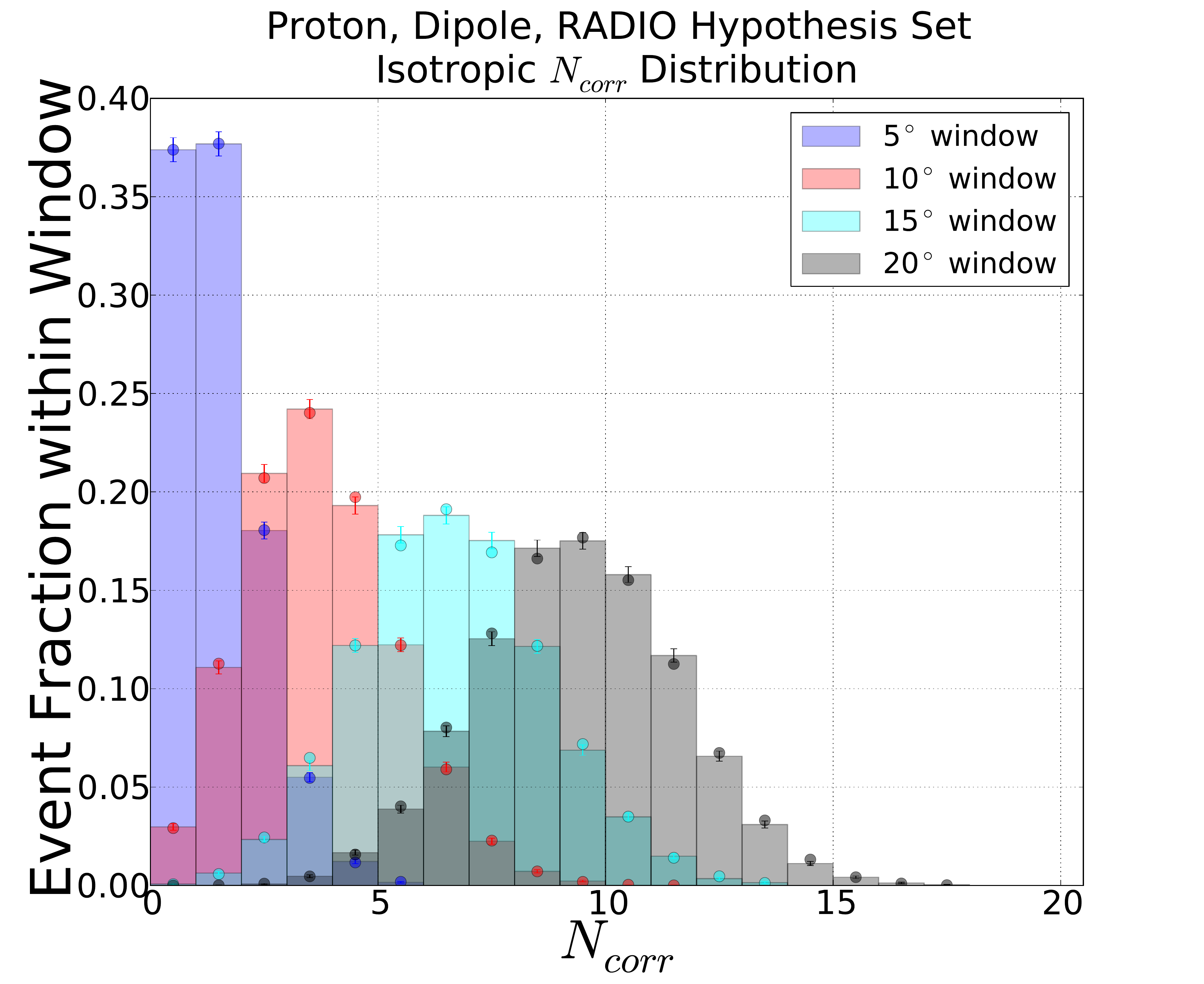}
\caption{Histogram of $N_{corr}$ for different angular windows for $10^{4}$ isotropic instances at $B_{\odot}=1.5\ \mu$G.
The error bars are Poisson error statistics on each histogram bin entry.
The dots show the probability values for a binomial distribution derived from the mean correlation fraction of each angular window histogram.}
\label{fig:p-d-r_ncorr_nsim0}
\end{figure}

Figure \ref{fig:p-d-r_ncorr_nsim0} shows a histogram of the correlation number for various angular windows for $10^{4}$ isotropic instances at $B_{\odot}=1.5\ \mu$G.
The error bars are calculated using Poisson statistics on each histogram bin entry.
The circles depict binomial distributions calculated according to $p=\overline{N}^{iso}_{corr}$ / $N_{tot}$ as described previously and are color-coded to their respective histograms.
KS tests performed between each $N^{iso}_{corr}$ distribution and its derived binomial distribution show clear agreement.
For small angular windows the isotropic correlation number distribution is well-described by a binomial distribution.
The binomial distribution approximation for small angular windows remains applicable even for parameter values and hypothesis sets distinct from truth.

\begin{figure}
\includegraphics[width=0.7\textwidth]{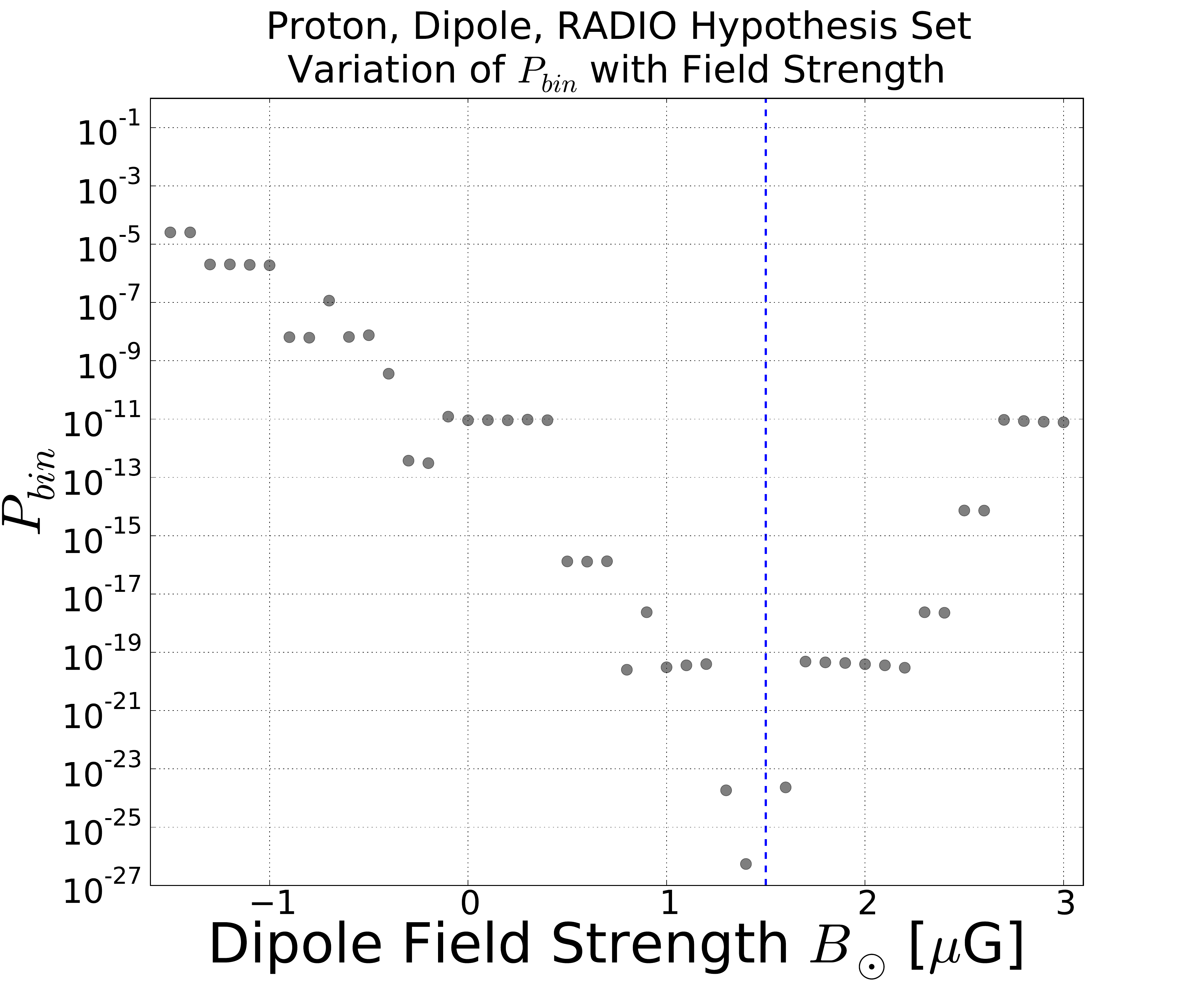}
\caption{Field scan of the truth scenario.
The truth parameter point is at $B_{\odot}=1.5\ \mu$G as indicated by the dashed blue line.}
\label{fig:p-d-r_Pbin}
\end{figure}

Figure \ref{fig:p-d-r_Pbin} shows $P_{bin}$ calculated from $N_{corr}^{DOI}$ according to Eq. \ref{eqn:binomial}.
The minimum $P_{bin}$ occurs at $B_{\odot}=1.4\ \mu$G due to statistical variation in the $\overline{N}^{iso}_{corr}$, however, the strongest correlation (all 20 events) is correctly observed in the adjacent field strength bin.
It is clear from these figures that both methods correctly identify the magnetic field parameters of the truth hypothesis set and that the hypothesis set is most self-consistent at those values.
As the field strength is detuned from the truth value the hypothesis set does not allow for complete correlation and becomes a self-\emph{in}consistent description of the cosmic ray observations.

We note that such extreme values of $P_{KS}$ and $P_{bin}$ arise from the contrived nature of the scenario.
Any realistic application of this method to real cosmic ray observations would lead to much larger values and smaller variations of these probability measures in accordance with the greater level of uncertainty in real datasets and knowledge of their environment.

\begin{figure}
  \centering
  \subfloat[]{\label{fig:p-d-rs_Pks}\includegraphics[width=0.5\textwidth]{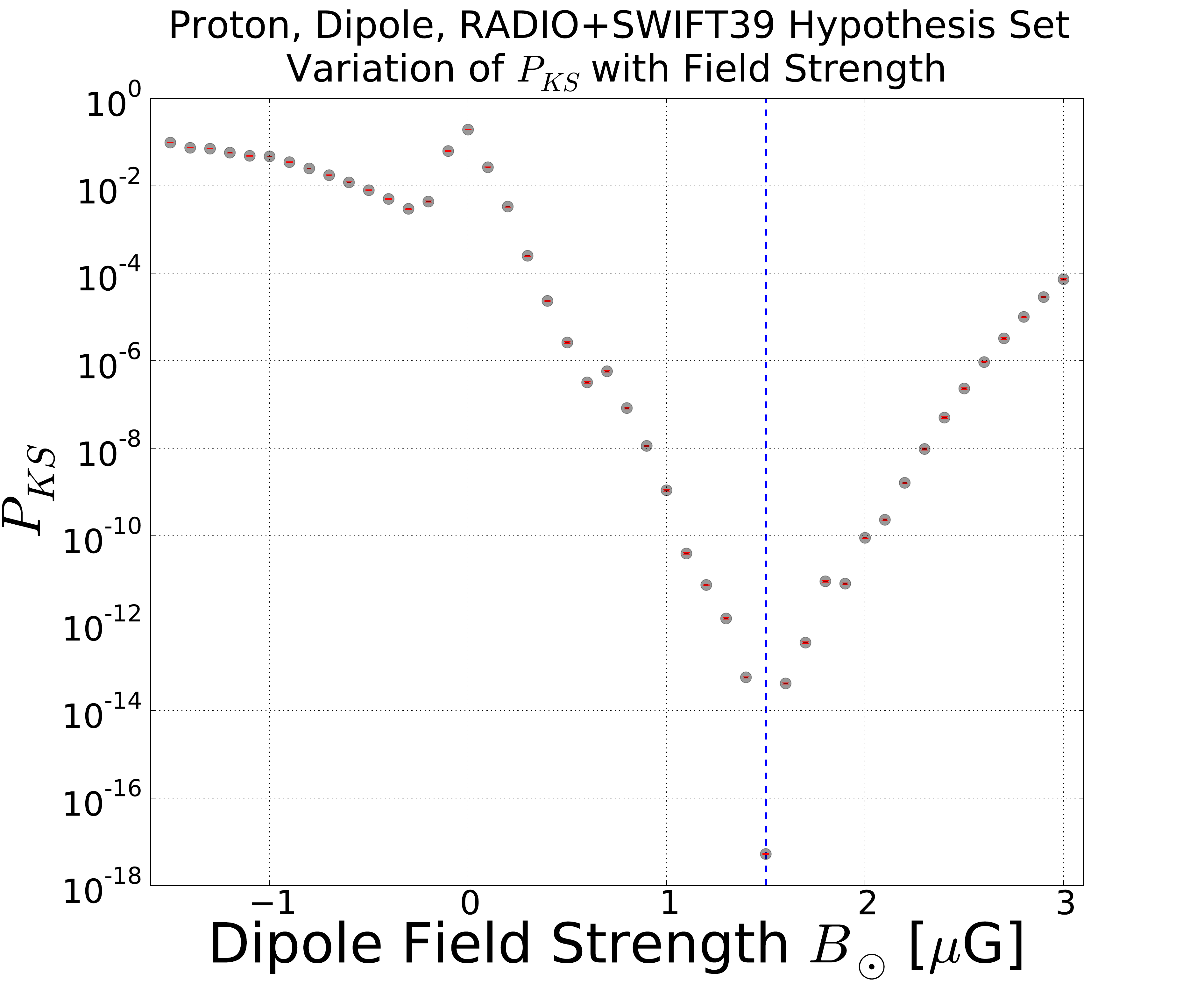}}
  \subfloat[]{\label{fig:p-d-i_Pks}\includegraphics[width=0.5\textwidth]{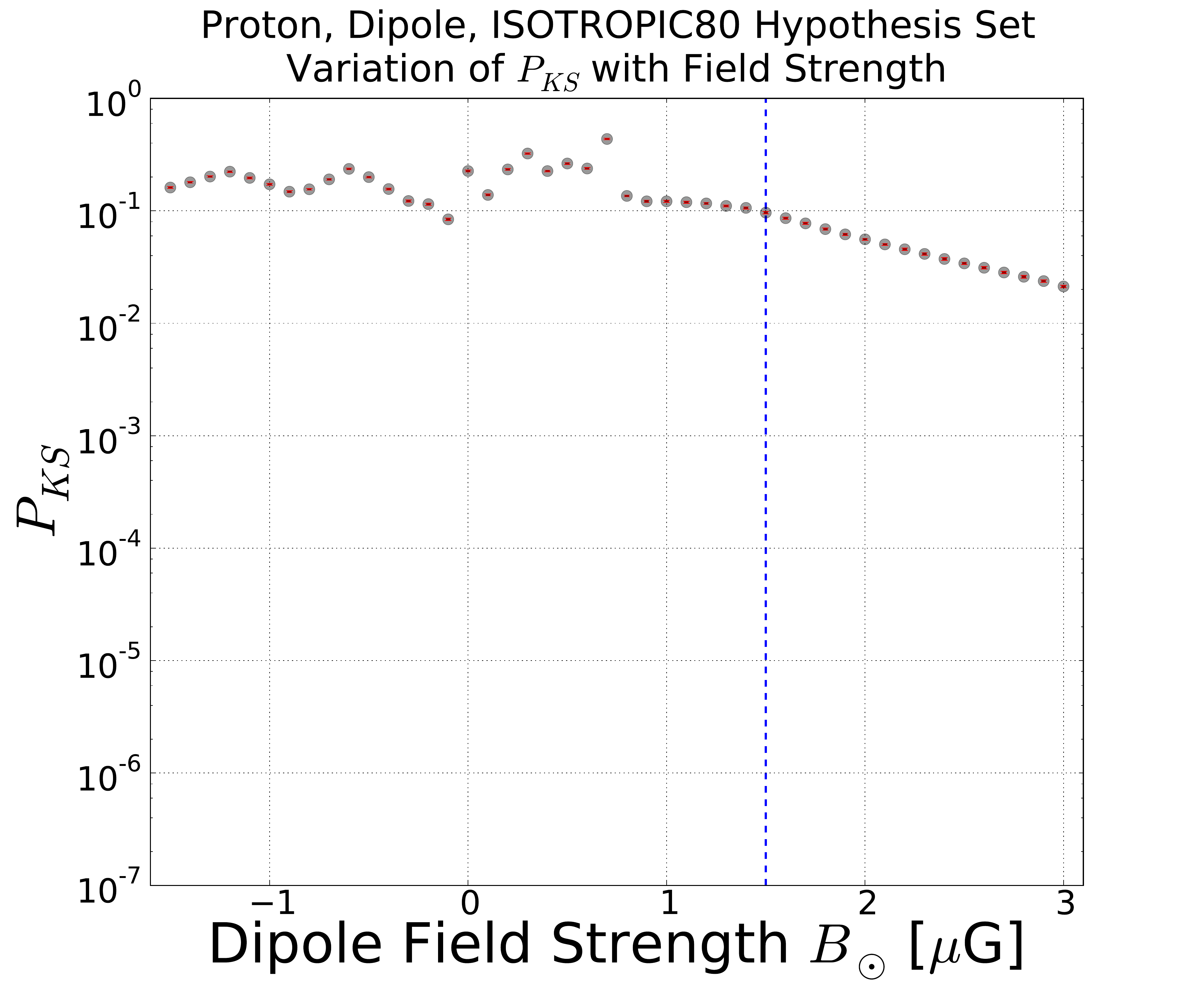}}
  \caption{Variation of hypothesis set components.
The field is the dipole model and the truth parameter point is at $B_{\odot}=1.5\ \mu$G as indicated by the dashed blue lines.
The source components are RADIO+SWIFT39 (left), and ISOTROPIC80 (right).
The shaded dots and red lines indicate the mean and spread of the $P_{KS}$ distribution from the 100 isotropic simulations.}
  \label{fig:vary_sources_pks}
\end{figure}

\begin{figure}
  \centering
  \subfloat[]{\label{fig:p-d-rs_Pbin}\includegraphics[width=0.5\textwidth]{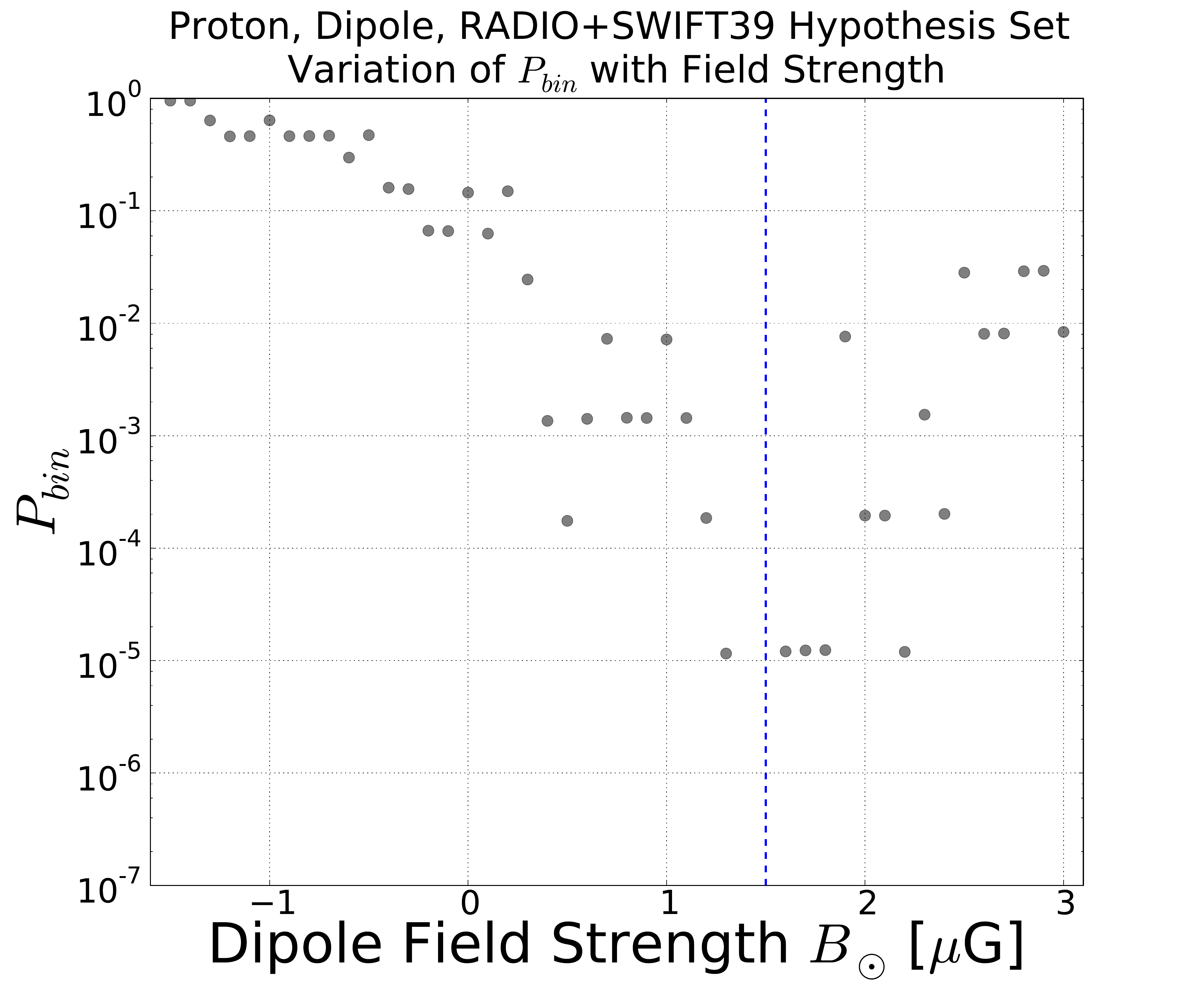}}
  \subfloat[]{\label{fig:p-d-i_Pbin}\includegraphics[width=0.5\textwidth]{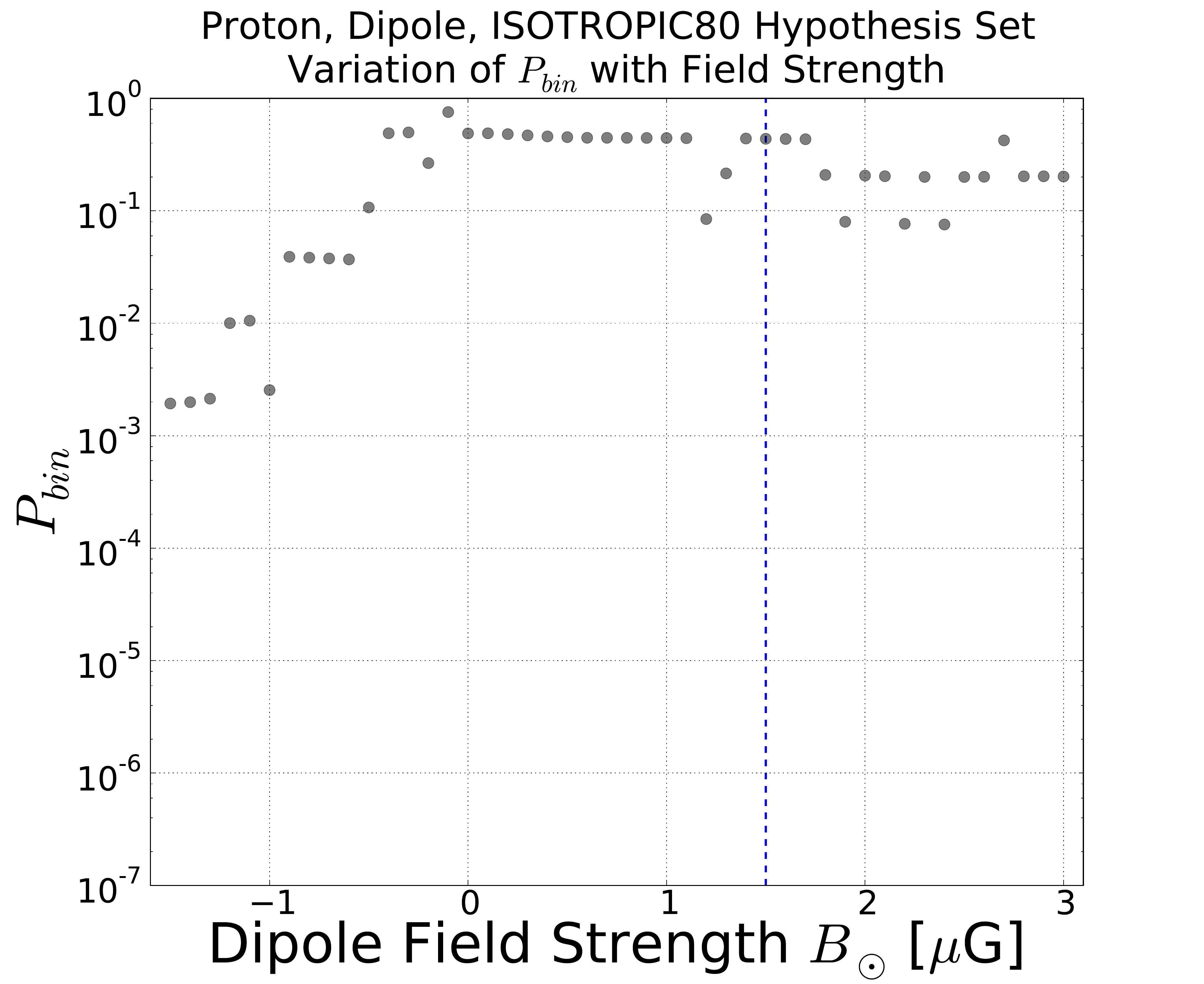}}
  \caption{Variation of hypothesis set components.
The field is the dipole model and the truth parameter point is at $B_{\odot}=1.5\ \mu$G as indicated by the dashed blue lines.
The source components are RADIO+SWIFT39 (left), and ISOTROPIC80 (right).}
  \label{fig:vary_sources_Pbin}
\end{figure}

We now begin to vary the components of the hypothesis sets.
Figures \ref{fig:vary_sources_pks} and \ref{fig:vary_sources_Pbin} show the variation of $P_{KS}$ and $P_{bin}$ for hypothesis sets incorporating RADIO+SWIFT39 and ISOTROPIC80 with a proton composition and the dipole model.
Due to the fact that some of the true sources are a subset of the RADIO+SWIFT39, $P_{KS}$ is minimized at a comparable value to that in Figure \ref{fig:p-d-r_Pks} since the $\Psi_{i}^{DOI}$ distribution is unchanged.
The DOI events are not distributed similarly to isotropy after being backtracked.
However, Figure \ref{fig:p-d-rs_Pbin} shows that source correlation is more consistent with the mean isotropic expectation.
At $B_{\odot}=1.5\ \mu$G, $N_{corr}^{DOI}=20$ but the larger number of source objects in this catalog increases the number of random correlations so that $P_{bin}$ is increased to $10^{-5}$.
Although such a large deviation from isotropy and high value of $N_{corr}^{DOI}$ could prompt further investigation, these results do not indicate a self-consistent description of the hypothesis set.

The hypothesis set incorporating the ISOTROPIC80 source distribution is also observed to not be self-consistent in Figures \ref{fig:p-d-i_Pks} and \ref{fig:p-d-i_Pbin}.
The $\Psi_{i}^{DOI}$ distribution is consistent with the isotropic expectation.
Although it begins to decrease at larger field strength it still lies at large values ($P_{KS}>10^{-2}$).
However, no correlation excess is observed in concert with a smaller $P_{KS}$.
This indicates that the backtracked DOI is not behaving in a manner similar to isotropy, but it is not correlating with the hypothesized source distribution, therefore ruling this hypothesis set self-inconsistent.

\begin{figure}
  \centering
  \subfloat[]{\label{fig:p-u-r_Pks}\includegraphics[width=0.33\textwidth]{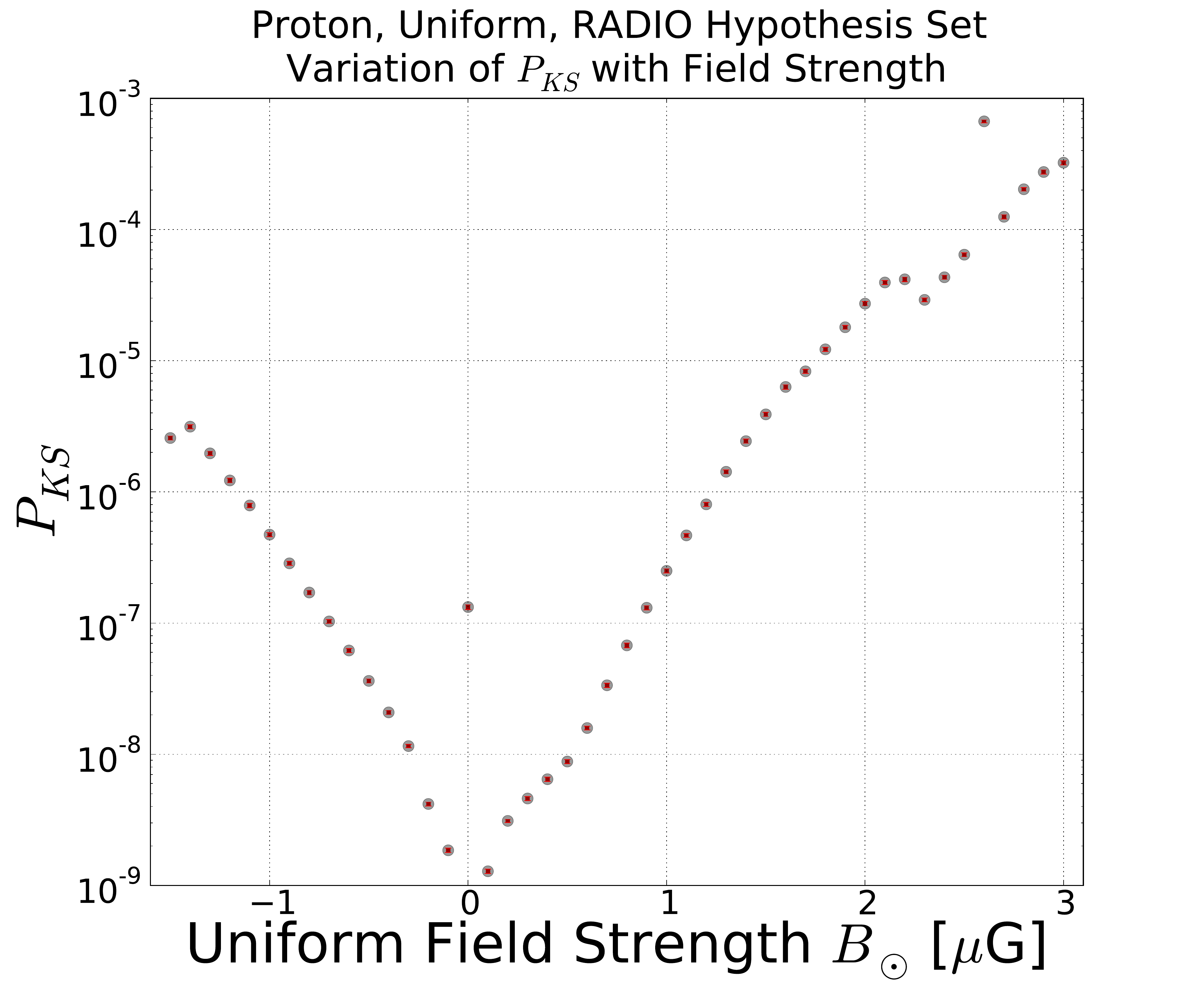}}
  \subfloat[]{\label{fig:p-u-rs_Pks}\includegraphics[width=0.33\textwidth]{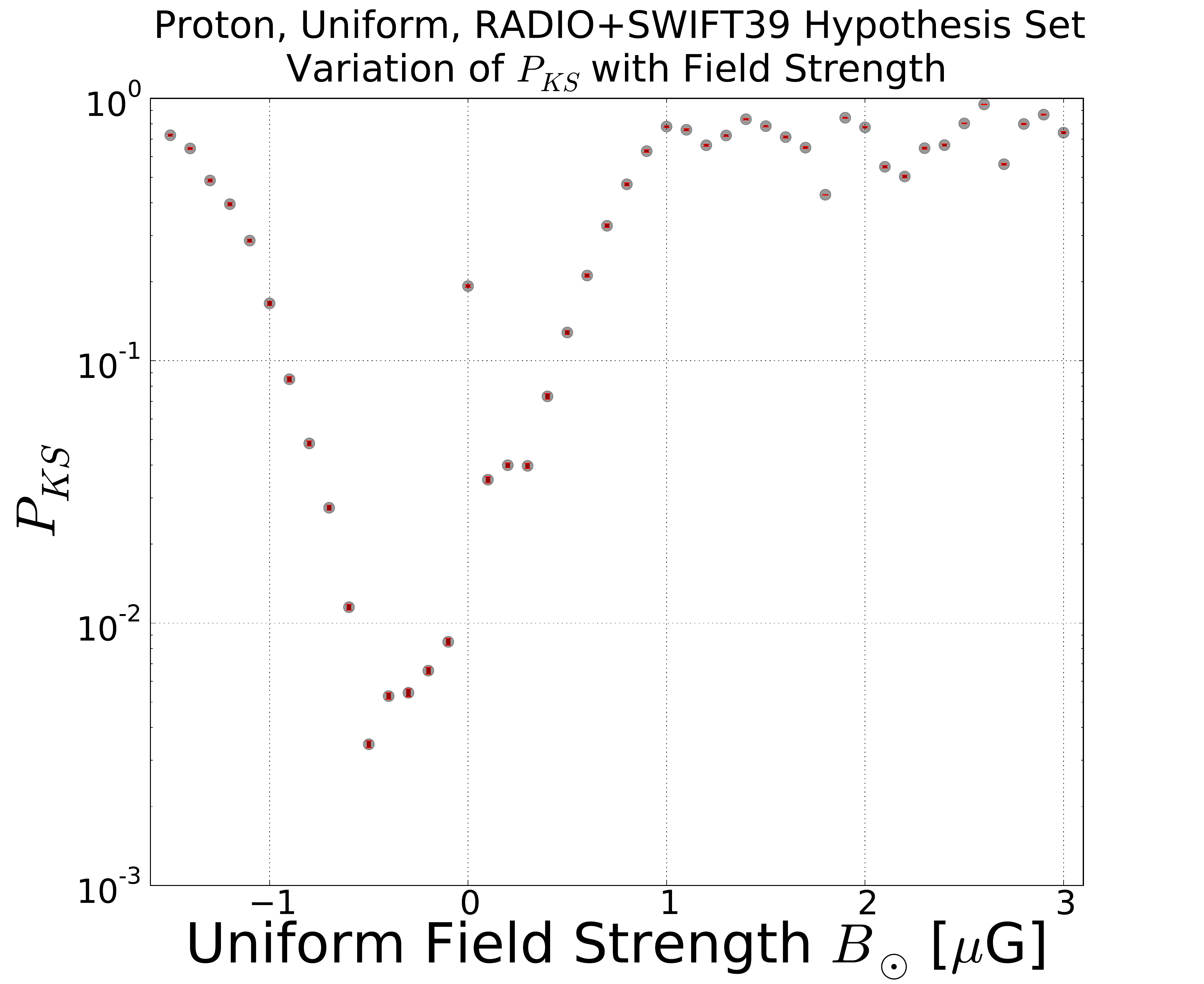}}
  \subfloat[]{\label{fig:p-u-i_Pks}\includegraphics[width=0.33\textwidth]{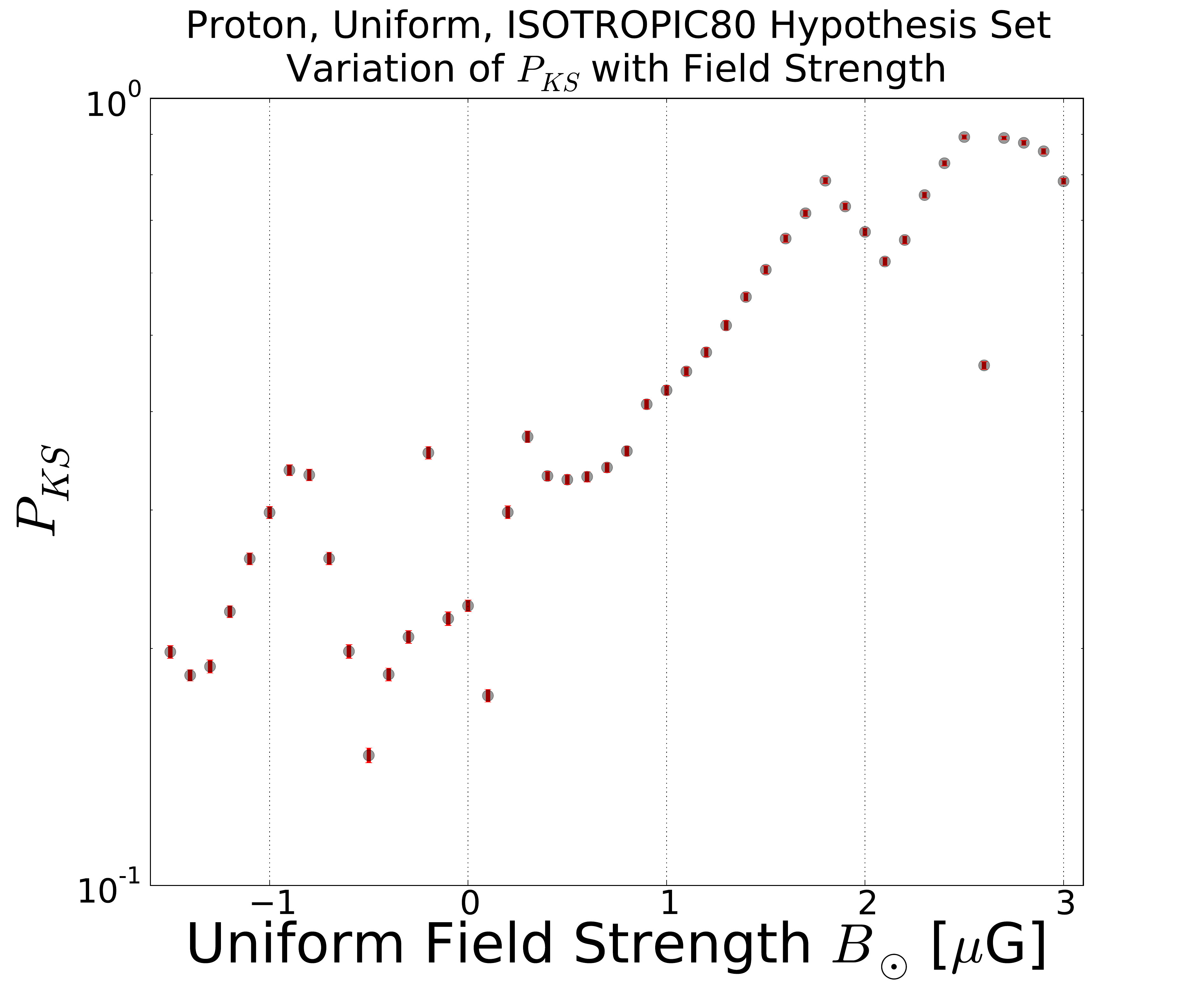}}
  \caption{Variation of hypothesis set components.
The field is the uniform model.
The source components are RADIO (left), RADIO+SWIFT39 (middle), and ISOTROPIC80 (right).
The shaded dots and red lines indicate the mean and spread of the $P_{KS}$ distribution from the 100 isotropic simulations.}
  \label{fig:vary_fields_pks}
\end{figure}

\begin{figure}
  \centering
  \subfloat[]{\label{fig:p-u-r_Pbin}\includegraphics[width=0.33\textwidth]{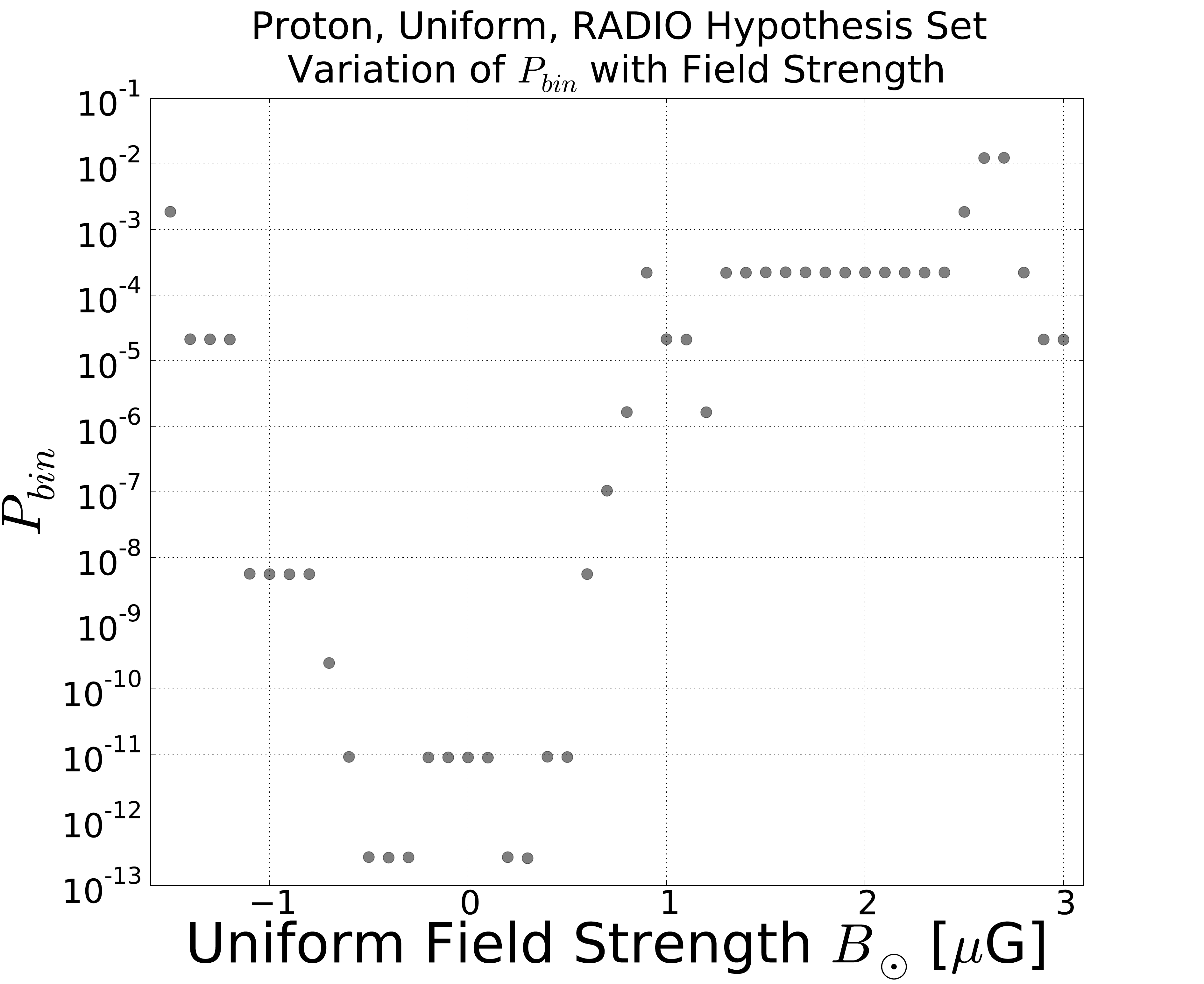}}
  \subfloat[]{\label{fig:p-u-rs_Pbin}\includegraphics[width=0.33\textwidth]{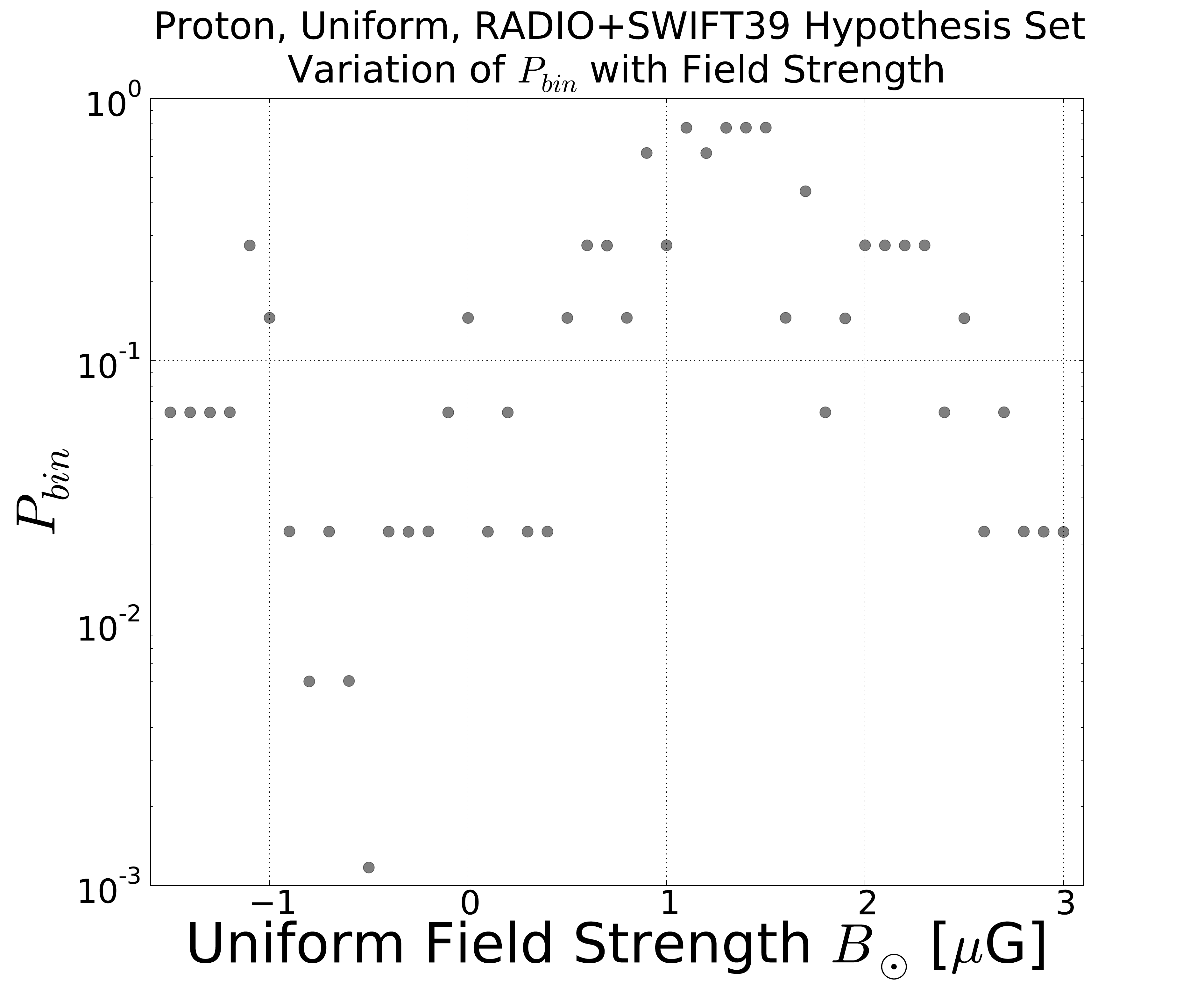}}
  \subfloat[]{\label{fig:p-u-i_Pbin}\includegraphics[width=0.33\textwidth]{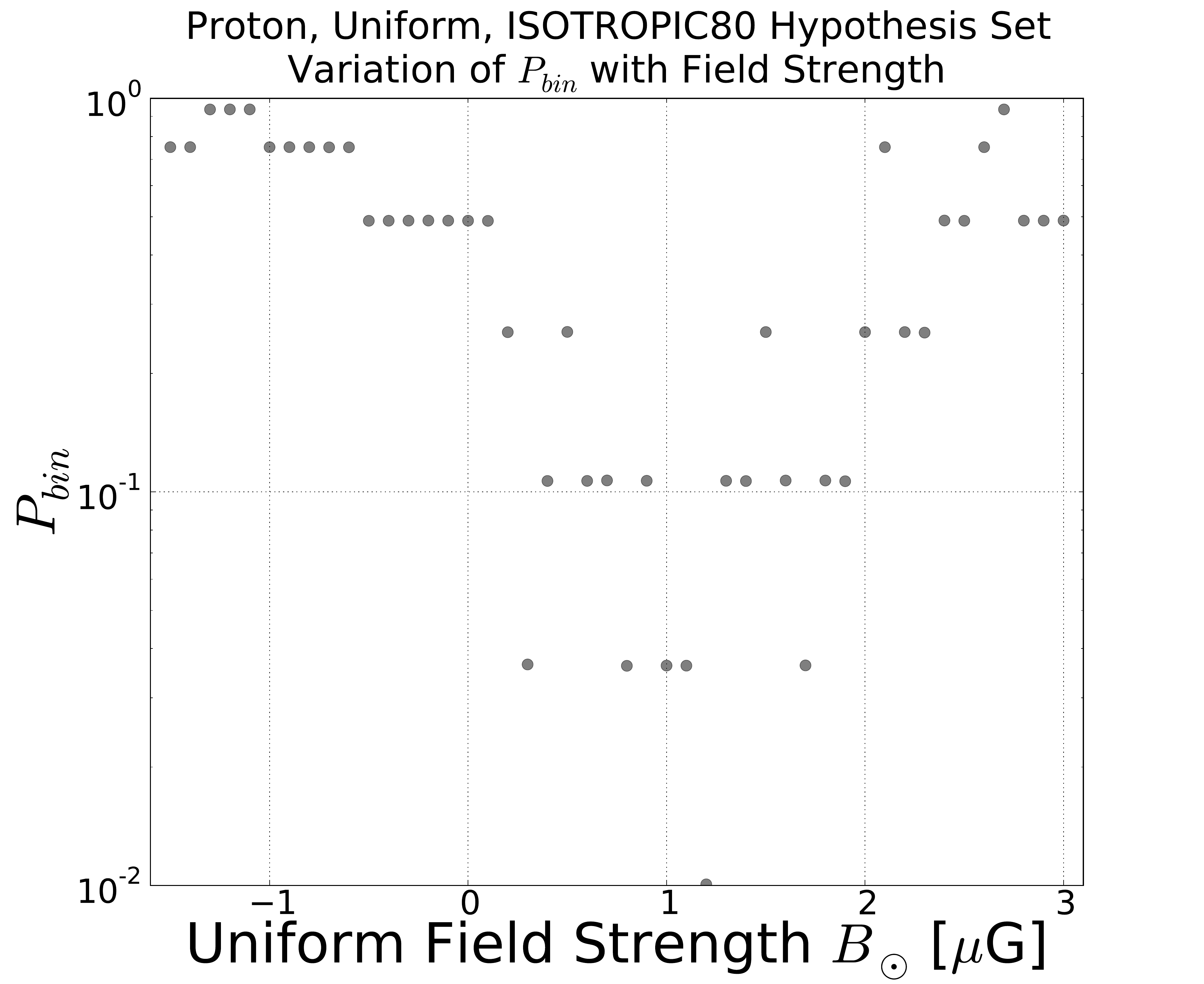}}
  \caption{Variation of hypothesis set components.
The field is the uniform model.
The source components are RADIO (left), RADIO+SWIFT39 (middle), and ISOTROPIC80 (right).}
  \label{fig:vary_fields_Pbin}
\end{figure}

Figures \ref{fig:vary_fields_pks} and \ref{fig:vary_fields_Pbin} show the variation of $P_{KS}$ and $N_{corr}$ versus the field strength for hypothesis sets incorporating the uniform field model.
These 3 hypothesis sets do not increase source correlations and are not self-consistent descriptions.
Indeed for the RADIO source hypothesis in Figure \ref{fig:p-u-r_Pbin}, while the ``observed'' ($B_{\odot}=0\ \mu$G) $P_{bin}$ is small, the presence of a nonzero field serves only to lower the correlation count which contradicts the explicit assumption that the hypothesis set is correct.

\subsection{Additional Scenarios}
\label{sec:validation:otherscenarios}
In addition to the simple scenario described previously, we have also investigated situations where the truth DOI UHECR properties and GMF are more consistent with observations and theoretical expectations proposed in the literature \cite{1997ApJ...479..290S,1999JHEP...08..022H,2003A&A...410....1P,2009JCAP...07..021J}.
The truth of two such scenarios are detailed in Table \ref{tbl:mock-truth-params}.
The VCV catalog listed in the table is the 12$^{th}$ Edition V\'{e}ron-Cetty and V\'{e}ron catalog of quasars and active nuclei \cite{2006yCat.7248....0V}.
The BSS$\_$A and ASS$\_$S field types are logarithmic spiral GMF models \cite{1999JHEP...08..022H}.
These scenarios are representative of the results found during the extensive validation process.
\begin{table}[htp]
\begin{center}
\caption{Truth Scenarios}
\begin{tabular}{|c|c|c|}
 \hline
                                    &   \bf{I}          &   \bf{II}          \\
 \hline
 \hline
 Composition                        &   Pure proton     &   Pure Iron        \\
 \hline
 Isotropic Fraction                 &   50\%             &   0\%             \\
 \hline
 Source Distribution                &   VCV Catalog ($z \leq 0.017$)     &   Swift 39 Month Catalog ($z \leq 0.021$)     \\
 \hline
 Field Type                         &   BSS$\_$A    &   ASS$\_$S     \\
 \hline
 Local Field Strength ($B_{\odot}$) &   $0.71\ \mu$G    &   $1.9\ \mu$G     \\
 \hline
 Planar Scale Height ($z_{1}$)      &   0.95 kpc       &   1.2 kpc        \\
 \hline
 Field Pitch Angle ($p$)            &   -9.6$\deg$ &   -13.1$\deg$  \\
 \hline
\end{tabular}
\label{tbl:mock-truth-params}
\end{center}
\end{table}

\begin{figure}
\includegraphics[width=0.7\textwidth]{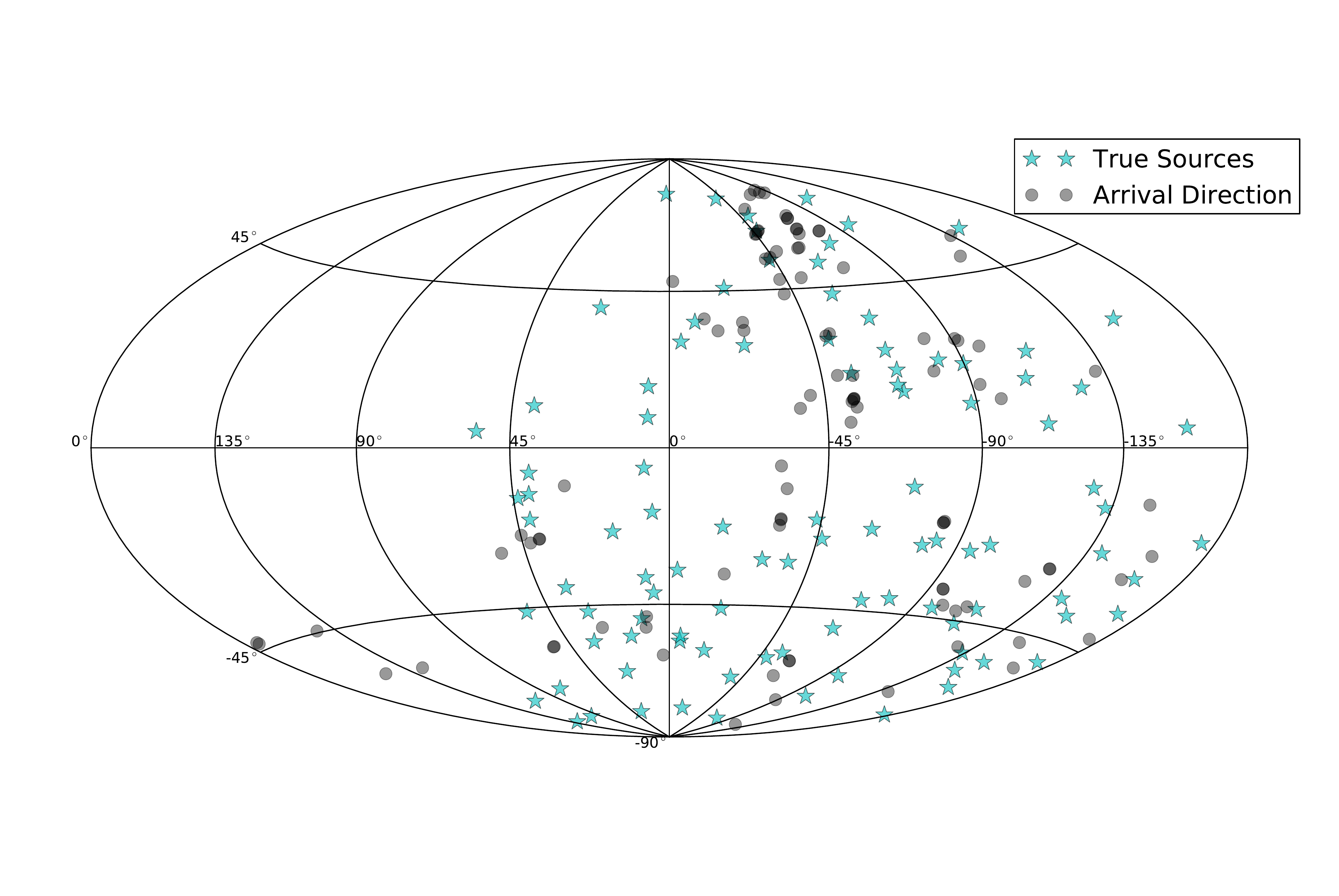}
\caption{Observed arrival directions (shaded black circles) and positions of their true sources (blue stars) of the events in the realistic Truth Scenario I.}
\label{fig:realistic_arrival}
\end{figure}

To make the validation more realistic the fraction of source to isotropic (background) events has also been varied\footnote{Any set of UHECR observations will contain events which come from sources not in a catalog and not directly from the source, e.g. GZK daughter particles.}.
The arrival directions of the mock events incorporate observational resolution\footnote{These scenarios use the detector resolution of the Pierre Auger Observatory \cite{2009NuPhS.190...20T}.} and the sky coverage matches that of the southern site of the Pierre Auger Observatory, as shown in Figure \ref{fig:realistic_arrival}.
During the hypothesis set scanning procedure, additional deflection due to an extragalactic turbulent magnetic field is applied to the post-backtracked arrival direction by sampling a gaussian smear function centered on the backtracked direction; propagation through extragalactic space is not performed by \textit{CRT}.

\begin{figure}
\includegraphics[width=0.65\textwidth]{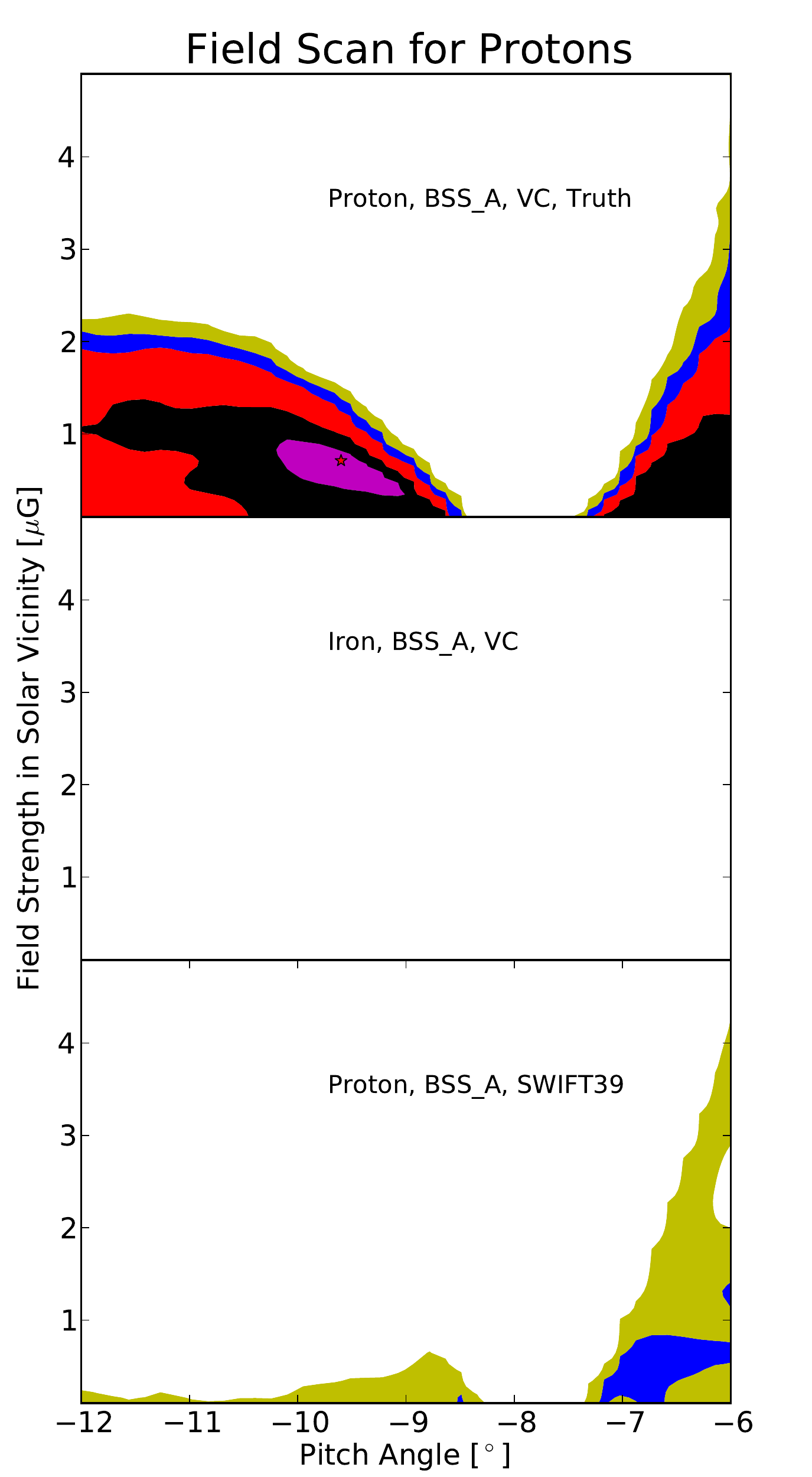}
\caption{Contour plot of $P_{KS}$ for a realistic scenario.
The subplots are labelled with their hypothesis set components.
The color scale of $P_{KS}$ is White, $P_{KS} \geq 10^{-4}$; Yellow, $10^{-4}  > P_{KS} \geq 10^{-5}$; Blue, $10^{-5}  > P_{KS} \geq 10^{-6}$; Red, $10^{-6}  > P_{KS} \geq 10^{-8}$; Black, $10^{-8}  > P_{KS} \geq 10^{-10}$; Magneta, $10^{-10}  > P_{KS}$.
The star in the top subplot indicates the truth hypothesis set parameters.}
\label{fig:truthlelmcat}
\end{figure}

The method behaves as expected near the truth values and does not produce false positives for all realistic validation scenarios.
Proton truth scenarios continue to show a dramatic decrease in the consistency between the DOI and the MCISOs, which is maximal in the magnetic field parameter space immediately surrounding the truth value.
Furthermore, as the field strength is increased the consistency decreases towards the truth value.
That is, as the field turns on, correlations with the source distribution increase more than would be expected from an isotropic sample.
Figure \ref{fig:truthlelmcat} shows contour plots for various hypothesis sets for the Truth Scenario I as functions of the local field strength and orientation.
The value of $P_{KS}$ at each point is smeared using a two dimensional gaussian kernel to minimize the effects of limited statistical samples.
The un-smeared surfaces do not show any important features not displayed in the smeared figures.
The truth realization is indicated as a star in the corresponding parameter space subplot.
Clear deviation from the isotropic expectation is observed; maximal deviation from isotropy occurs at the same parameter point as maximal source correlation for the truth hypothesis set.

The presence of a turbulent Galactic magnetic field component in addition to the regular component may have a significant effect on the identification of hypothesis set self-consistency.
In a separate study \cite{2009arXiv0906.2347T} it was found that magnetic turbulence typically isotropizes cosmic rays.
Compared to the $P_{KS}$ value for a GMF with no turbulence, $P_{KS}$ for a turbulent GMF is typically larger.
This can hinder identification of significant source correlation and self-consistency since simultaneous decreases in $P_{KS}$ and $P_{bin}$ may not be observed.

There are some of the limitations of the FSM.
Specifically,  all scenarios incorporating an iron composition are found to be consistent with isotropy.
This is to be expected since the arrival directions of UHECR iron primaries have been effectively isotropized with respect to the hypothesized source distributions by the GMF for all but the smallest $B_{\odot}$\footnote{As stated above the FSM relies on the existence of an underlying anisotropy in the observed arrival directions of UHECRs.}.
Magnetic parameter space would have to be so finely gridded to see meaningful variation in $P_{KS}$ and $P_{bin}$ as to be computationally limited.

\section{Summary}
\label{sec:fieldscan-summary}
A method for determining the self-consistency of GMF and UHECR property hypothesis sets based on cosmic rays observations has been presented and validated using simple and realistic scenarios.
Two complementary procedures are utilized which determine the correlation excess beyond the isotropic expectation as well as the behavior of the entirety of backtracked cosmic ray observations relative to that of the same isotropic simulations.
The FSM finds regions of magnetic field parameter space within a given overall hypothesis set which are maximally inconsistent with an isotropic assumption.
No false positives are found for the tested scenarios\footnote{Only regions near a truth scan point show a significant decrease in the consistency with isotropy.}.
Thus, parameter space of a hypothesis set where an already existing anisotropy is neither maintained nor increased can be efficiently eliminated from consideration as theories describing the UHECR observations.

The FSM does have limitations.
The method depends on finding regions of parameter space where anisotropic observations remain inconsistent with isotropic observations even after backtracking.
As such the method cannot be used on UHECR observations without an already existing observed anisotropy or with a composition hypothesis dominated by high Z UHECR.

\bibliographystyle{apsrev4-1}
\bibliography{document}

\end{document}